\documentclass[10pt,aps,showpacs,preprintnumbers,twocolumn,amsmath,amssymb,pre,floatfix,superscriptaddress,nofootinbib]{revtex4-1}

\usepackage{graphicx}

\usepackage[usenames,dvipsnames]{color}

\usepackage[caption=false]{subfig}

\usepackage{ifpdf}

\ifpdf
\usepackage[pdftex]{hyperref}
\else
\usepackage[hypertex,ps2pdf,dvips,colorlinks,bookmarks=false,urlcolor=blue,citecolor=NavyBlue,linkcolor=OliveGreen]{hyperref}
\fi

\def\be{\begin{equation}}
\def\ee{\end{equation}}
\def\bes{\begin{subequations}}
\def\esu{\end{subequations}}
\def\erf{\eqref}

\newcommand{\ud}          {\mathrm d}
\newcommand\eps           {\varepsilon}
\newcommand\w           {\omega}

\newcommand\p             {\partial}

\renewcommand\th          {\theta}
\newcommand\kb            {k_\text{B}}
\newcommand\pt            {p_\text{T}}

\renewcommand\pl            {\lambda_\text{l}}
\newcommand\pr           {\lambda_\text{r}}

\newcommand\kl            {k_\text{l}}
\newcommand\kr           {k_\text{r}}

\newcommand\sG 		{sine--Gordon }

\newcommand{\vev}[1]{\left\langle #1 \right\rangle}
\newcommand{\vevn}[1]{\big\langle #1 \big\rangle}

\begin{document}

\title{
Finite temperature dynamics in gapped 1D models\\ in the sine--Gordon family
}

\author{M. Kormos}
\affiliation{Department of Theoretical Physics, Institute of Physics, Budapest University of Technology and Economics, M\H uegyetem rkp. 3., H-1111 Budapest, Hungary}
\affiliation{MTA-BME Quantum Dynamics and Correlations Research Group, Budapest University of Technology and Economics, M\H uegyetem rkp. 3., H-1111 Budapest, Hungary}
\author{D. V\"or\"os}
\affiliation{Department of Theoretical Physics, Institute of Physics, Budapest University of Technology and Economics, M\H uegyetem rkp. 3., H-1111 Budapest, Hungary}
\affiliation{Institute for Solid State Physics and Optics, Wigner Research Centre for Physics, H-1525 Budapest, P.O. Box 49, Hungary}
\author{G. Zar\'and}
\affiliation{Department of Theoretical Physics, Institute of Physics, Budapest University of Technology and Economics, M\H uegyetem rkp. 3., H-1111 Budapest, Hungary}
\affiliation{MTA-BME Quantum Dynamics and Correlations Research Group, Budapest University of Technology and Economics, M\H uegyetem rkp. 3., H-1111 Budapest, Hungary}


\date{\today}

\begin{abstract}

The sine--Gordon model appears as the low-energy effective field theory of various one-dimensional gapped quantum systems. 
Here we investigate the dynamics of generic, non-integrable systems belonging to the sine--Gordon family at finite temperature within the semiclassical approach. Focusing on time scales where the effect of nontrivial quasiparticle scatterings becomes relevant, we obtain universal results for the long-time behavior of dynamical correlation functions. We find that correlation functions of vertex operators behave neither ballistically nor diffusively but follow a stretched exponential decay in time. We also study the full counting statistics of the topological current and find that distribution of the transferred charge is non-Gaussian with its cumulants scaling non-uniformly in time.

\end{abstract}

\maketitle

\section{Introduction}

Dynamical correlation functions are among the most important observables in quantum many-body systems, due to their direct connection with experimentally measurable quantities such as response functions. In real experiments, the effects of finite temperature are almost always important, and must be taken into account. However, real time correlations at finite temperature are extremely hard to access. Perturbation theory and Monte Carlo methods work mostly in imaginary time, and the analytic continuation to real time is nontrivial, especially in the low-frequency limit. A possible route is to study field theories emerging as low-energy effective descriptions of microscopic systems.

In this paper we focus on one of the most important models of this kind, the sine--Gordon (sG) quantum
field theory 
defined by the action
\be
S = \frac{c}{16\pi}\int \! \ud x\ud t \left[ \frac{1}{c^2} (\p_t\Phi)^2  -  (\p_x\Phi)^2 + g^2 \cos(\gamma\Phi)\right]\,,
\label{eq:action}
\ee
where $\Phi(x,t)$ is a real scalar field and $c$ is the speed of light. The sG model
provides the effective low-energy description of many one-dimensional gapped systems. Within the bosonization framework, the long-wavelength behavior of Luttinger liquids in the presence of a gap-opening perturbation is captured by the sG model \cite{Giamarchibook,Gogolin}. Prominent examples are given by the anisotropic Heisenberg spin chain in a staggered magnetic field \cite{Essler2005}, spin ladders \cite{Gogolin}, cold atoms in an optical lattice \cite{Buchler2003,Cazalilla_bosonizing2003}, and coupled cold atomic quasi-condensates \cite{Gritsev2007a,Schweigler2017}. 

The sG model is also a paradigmatic example of an integrable field theory \cite{Zamolodchikov1979,giuseppebook}. Integrability provides some special tools for computing correlation functions. One direction is based on the exact analytic expressions of form factors, matrix elements of local operators, that have been derived in this model. These can be used, in principle, in a spectral expansion of correlation functions. At finite temperature, the resulting form factor series, or linked cluster expansion, is plagued with singularities whose regularization requires a substantial amount of work \cite{Essler2009,Takacs2010,Szecsenyi2012}. Depending on the correlator, a partial resummation of the series may be necessary which is again a highly nontrivial task \cite{Essler2009}.

\begin{figure*}[tb]
\includegraphics[width=0.7\textwidth]{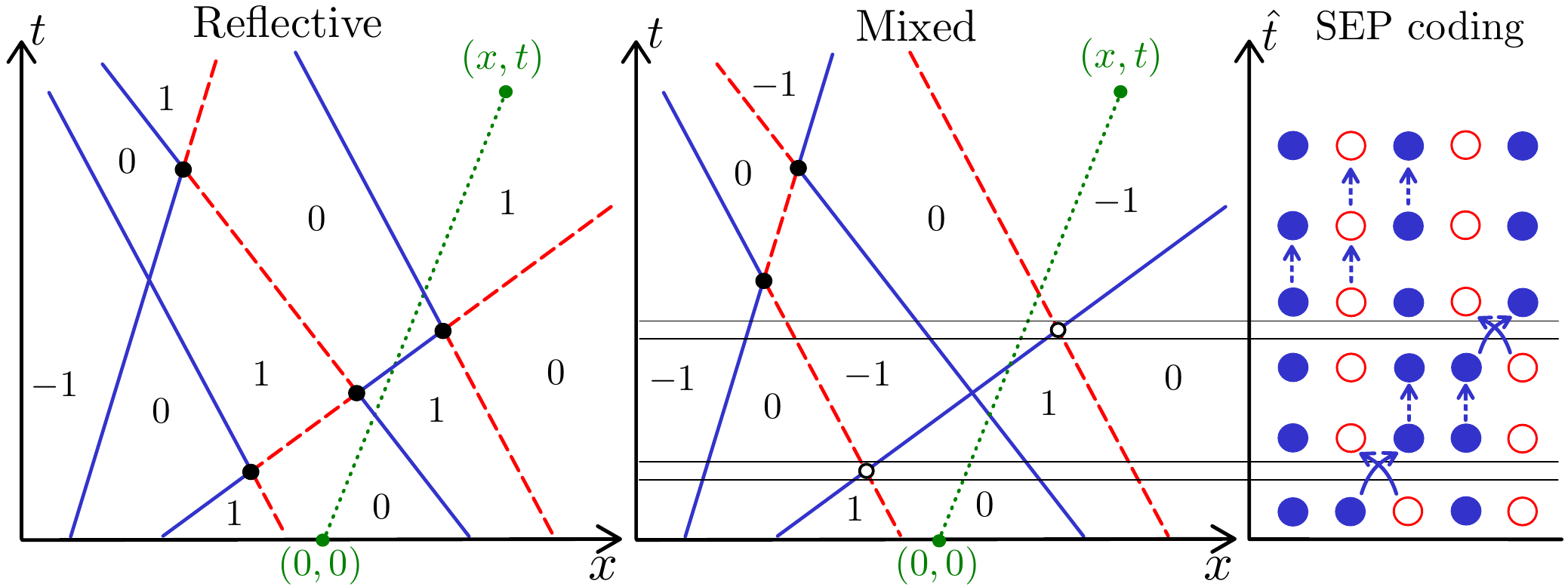}
\caption{Space-time diagrams of semiclassical histories. Kinks (solitons) travel along classical straight trajectories, while 
their topological charges follow zig-zag trajectories due to nontrivial collisions of solitons (blue) and antisolitons (red dashed lines). 
In the domains bordered by kink trajectories, the field sits in one of the vacua, $\Phi(x,t)= Q\,2\pi/\gamma,$ where the integer charges of the domains, $Q$ are displayed in the figure. {\it Left:} for perfectly reflective collisions, the spatial sequence of the domain charges is preserved. 
{\it Right:} the generic case with both reflective and transmissive collisions, and approximate mapping to the 
Simple Symmetric Exclusion Process (SEP).
 \label{fig:domains}}
\end{figure*}

Another possible route is provided by the framework of the recently developed generalized hydrodynamics (GHD) that describes the dynamics of integrable systems on hydrodynamic spatial and time scales, i.e., at the Euler scale based on the transport of the infinitely many conserved quantities \cite{Bertini2016,Castro-Alvaredo2016a,Doyon2019a}. These conserved charges naturally leave their mark on the dynamics, rendering the transport ballistic \cite{Prosen2013} or anomalous \cite{Ilievski2018,Ilievski2021}, in contrast to the generic diffusive behavior \cite{Medenjak2017} (for a recent review, see \cite{Bertini2021}). Expressions for correlation functions within the GHD formalism have also been derived \cite{Doyon2017f,Doyon2019b}.

However, the role of integrability is somewhat unclear in the applications of the sG model as an effective field theory for \emph{real} systems. Real systems are never perfectly integrable: a short distance cutoff spoiling integrability as well as irrelevant  terms neglected 
in the sG model are almost always  present. 
 Real systems  therefore lack the infinitely many conserved quantities that make the dynamics of integrable systems special. 
However, they do possess certain generic features of the sG  model such as the existence of a phase variable and corresponding 
soliton-like topological excitations. In this work, we shall refer to these models as belonging to the  sG \emph{family}.

For these reasons, to understand realistic models, integrability is most likely  inessential, 
which allows us  to take a less rigorous but more intuitive approach to capture  universal features of generic, non-integrable systems in the sG family. 
Our approach is based on the so-called semiclassical method proposed by Sachdev and collaborators \cite{Sachdev1996,Sachdev1997,Damle2005}. The essence of this method is the observation that at temperatures much lower than the gap, the system can be described in terms of a dilute gas of thermally excited massive quasiparticles that move along classical trajectories.\footnote{Note that the term ``semiclassical'' does not refer to the $\hbar\to0$ limit but rather to a low density limit.} In one spatial dimension, they will inevitably suffer collisions, which must be treated using the quantum mechanical scattering matrix. 
In the case of the sG model, these quasiparticles are solitons and antisolitons, possessing topological charges, 
and colliding with each other (see Fig.~\ref{fig:domains}).

In the original semiclassical approach, the scattering matrix is taken to be the ``universal'' low energy limit of the two-particle S-matrix corresponding to perfect reflections in terms of the internal quantum numbers (e.g. charges) of the quasiparticles (left panel in Fig.~\ref{fig:domains}). The method has been applied to the transverse field Ising chain \cite{Sachdev1996}, the $O(3)$ nonlinear sigma model \cite{Sachdev1997}, the Potts model \cite{Rapp2006} and the sine--Gordon model \cite{Damle2005}. It was later generalized to the study of out of equilibrium dynamics after quantum quenches \cite{Evangelisti2013,Kormos2015,Bertini2019}, and an improved version of the method, the so-called hybrid semiclassical approach has been developed to go beyond the purely reflective collisions and to follow the time evolution of the ``charge'' part of the wave function fully quantum mechanically \cite{Moca2016,Werner2019}.

For the sG family, analytic results have been derived for the finite temperature 
dynamical correlation function of vertex operators \cite{Damle2005},
\be
\label{eq:corrfn}
C_\eta(x,t)=\big\langle e^{i\eta \Phi(x,t)}e^{-i\eta \Phi(0,0)}\big\rangle_T
\,,
\ee
using the original semiclassical approach based on purely reflective collisions. These correlators as 
well as those of the topological charge density have been shown to exhibit {\em diffusive} $\sim1/\sqrt{t}$
behavior for large times. 

The main question of our work is, how {robust  these results are} against changing the scatterings by allowing transmissions. 
It is quite clear that the long time limit and the limit of purely reflective scattering do not 
commute: no matter how suppressed  non-reflective processes, after a long enough time their effects accumulate, which
is expected to change the qualitative behavior. In order to answer this question, we use a somewhat simpler approach than the hybrid semiclassical method. Instead of following the fate of the coherent superpositions created by the S-matrices in the charge sector, we turn the scattering amplitudes into probabilities, reminiscent of a master equation approach. We find that while the charge density correlator remains diffusive, the correlation functions of vertex operators change qualitatively, they asymptotically decay in a {\em stretched exponential} manner.

The paper is organized as follows. In Sec. \ref{sec:setup}, we set the stage by introducing the sine--Gordon model and discussing the details of the semiclassical approach. Sec. \ref{sec:corr} is dedicated to the study of the vertex operator correlation functions. After re-deriving the analytic results in the perfectly transmissive and reflective cases, we derive the long-time asymptotic behavior by mapping the charge dynamics for weak transmissions to the simple symmetric exclusion process. The results of our numerical semiclassical simulations are also presented here. In Sec. \ref{sec:PQ} we turn to the full counting statistics of the integrated topological current, while in Sec. \ref{sec:qq} we discuss the correlation functions of the topological charge density. We provide our conclusions and outlook in Sec. \ref{sec:concl}.

\section{Setup}
\label{sec:setup}

The action of the \sG model is given in Eq. \erf{eq:action}. We work in the $\hbar=\kb=1$ convention so both $\Phi$ and the coupling constant $\gamma$ are dimensionless. 
%
Here we focus on the so-called repulsive regime of the model,
$1/\sqrt{2}\le\gamma<1,$
where the cosine term is relevant, the model is gapped, and the spectrum is built of multi-particle scattering states of massive kinks, i.e. solitons and antisolitons of topological charge $q=\pm1$ that interpolate between neighboring minima of the cosine potential.\footnote{For $0<\gamma<1/\sqrt2,$ neutral bound states of kinks called breather particles are also present in the spectrum.} The kinks have relativistic dispersion relation, $\eps(p)=\sqrt{p^2c^2+M^2c^4},$ 
where their mass, $M$, can be explicitly expressed in terms of $g$ and $\gamma$ \cite{Zamolodchikov1995}. The energy and momentum of the incoming and outgoing particles can be conveniently parameterized in terms of the relativistic rapidity $\th$ as $E=Mc^2\cosh\th$ and $p=Mc\sinh\th,$  respectively.

The sG model is integrable, therefore multi-kink scattering processes factorize into two-particle scatterings
with an exactly known  two-particle S-matrix~\cite{Zamolodchikov1979}. Due to topological charge conservation, 
kinks of the same topological charge only suffer a phase shift upon collision, while scattering in the $q=0$ (soliton-antisoliton) channel is nontrivial: 
a colliding soliton and antisoliton can get reflected or transmitted. The 2-particle S-matrix is thus given by a four by four block-diagonal matrix in the 2-kink basis $|++\rangle,|+-\rangle,|-+\rangle,|--\rangle$ labeled by the topological charges of the kinks:
\begin{align}
\label{eq:S}
&\quad\;|++\rangle\;|+-\rangle\;|-+\rangle\;|--\rangle\\
S = 
&
\left(
\begin{tabular}{c|cc|c}
$S_0$&\phantom{00}0\phantom{00}&\phantom{00}0\phantom{00}&\phantom{00}0\phantom{00}\\
\hline
\phantom{00}0\phantom{00}&$S_R$&$S_T$&0\\
0&$S_T$&$S_R$&0\\
\hline
0&0&0&$S_0$\\
\end{tabular}
\right)
\begin{matrix}
|++\rangle\\
|+-\rangle\\
|-+\rangle\\
|--\rangle
\end{matrix}\;,
\end{align}
where, due to relativistic invariance, all entries depend only on the relative rapidity $\theta=\theta_1-\theta_2$ of the two incoming kinks. Here $S_\text{T}(\theta)$ is the amplitude of transmission and $S_\text{R}(\theta)$ is the amplitude of reflection. The term $S_0$ accounts for the phase picked up by the wave function upon scattering of two kinks of the same charge.
The transmission and reflection factors are given by
\begin{align}
S_\text{T} (\th)&= \frac{\sinh\frac{\theta}\xi}{\sinh\frac{i\pi-\theta}\xi}\,S_0(\theta)\,,\label{eq:ST}\\
S_\text{R} (\th)&= i\frac{\sin\frac{\pi}\xi}{\sinh\frac{i\pi-\theta}\xi}\,S_0(\theta)\,,
\end{align}
where
$
\xi
= \frac{\gamma^2}{1-\gamma^2}.
$
%
The probabilities of transmission and reflection are given by the modulus squares of the amplitudes which satisfy $|S_\text{T}|^2+|S_\text{R}|^2=1.$ At $\gamma=1/\sqrt{2}$ ($\xi=1$), the reflection probability is exactly zero for any $\theta$. This is the so-called free fermion point, where the model can be mapped to the theory of free Dirac fermions \cite{Luther-Emery1974,Coleman1975}. For small $\theta,$ they behave as $|S_\text{T}|^2\propto \theta^2$ and $|S_\text{R}|^2\propto1-\theta^2$, thus at small incoming momenta, the scattering of kinks of opposite charges is almost purely reflective, which is at the heart of the semiclassical approach.

As discussed in the Introduction, we aim to describe generic systems in the sG family, where integrability is (weakly) broken.
 We assume the existence of two species of long-lived massive quasiparticles of opposite topological charge and with properties of 
 the sG kinks. While the S-matrix may be slightly modified by  integrability breaking terms, certain  properties of the S-matrix are 
 the same as in the sG model. Quasiparticles with vanishing momenta, e.g., can be shown to scatter reflectively on each other, 
just as in the sG case~\cite{Damle2005}.
 Since  our main results and conclusions will not depend on the exact form of the S-matrix,  for the sake of simplicity and concreteness, we shall 
 use the sG S-matrix \erf{eq:S} in what follows.

The semiclassical approach is based on the observation that at low temperatures, both the density and the velocities of the
 kinks are suppressed and the thermal state, in the absence of a chemical potential, corresponds to a dilute 
 neutral gas of slow kinks. In a Keldysh path integral approach, the main contribution to a dynamical 
 correlation function is given by classical histories corresponding to kinks moving along straight trajectories between rare collisions 
 (see Fig.~\ref{fig:domains}).\footnote{Note that due to the diluteness of the gas, multi-particle scatterings can be neglected even away from integrability.} 
 In the semiclassical approach, we compute thermal expectation values by averaging over these semiclassical histories 
 in a Monte Carlo fashion. Each history is specified by the initial configuration of the kinks and the outcomes 
 of the scattering events. We discuss these two aspects in turn.

The initial states correspond to a spatially uniform thermal gas of kinks of randomly assigned topological charges ($q=\pm1$ with probability $1/2$) with a classical (but relativistic) velocity distribution. The density of kinks is thus given by
\be
\rho = \int\frac{\ud p}{2\pi}f(p) = \int\frac{\ud p}{2\pi} e^{-\sqrt{p^2c^2+M^2c^4}/T}\,.
\ee
At low temperatures, 
$T\ll Mc^2,$
the density is well approximated by the nonrelativistic formula
%
$
\rho\approx \sqrt{MT/(2\pi)}e^{-Mc^2/T}.
$
%
The normalized velocity distribution for $v\in(-c,c)$ is given by 
\be
\label{fv}
f(v) = \frac{1}{2\pi\rho}\frac{\ud p}{\ud v} f(p(v)) =  \Theta(c-|v|)\frac{M}{2\pi\rho}\frac{ e^{-\frac{Mc^2/T}{\sqrt{1-v^2/c^2}}} }{(1-v^2/c^2)^{3/2}} \,,
\ee
where the Heaviside theta function expresses that the velocity satisfies $|v|<c$. For $T\ll M,$ we recover the 
Maxwell--Boltzmann distribution $f(v)\approx \sqrt{M/(2\pi T)}e^{-\frac{Mv^2}{2T}}.$
For later use, we introduce the characteristic collision time $\tau$, as the ratio of the mean interparticle distance and the average velocity,
\be
\label{eq:tau}
\tau^{-1}\! \equiv \frac{\bar v}{1/\rho} = \rho \int_{-c}^c \ud v f(v) |v| 
=
\frac{T}\pi \,e^{-Mc^2/T},
\ee
where we used $v=\ud \eps(p)/\ud p.$

Having discussed  the initial state, let us now turn to the collisions. Our semiclassical approach is in the spirit of  Pauli's master equation: 
we neglect interference between consecutive scatterings, and we  treat collisions as a classical process. At each collision of two kinks of opposite topological charge, a reflection or transmission is realized with  probabilities set by the modulus 
square of the exact S-matrix amplitudes, which depend only on the incoming momenta of the particles. The collision 
history thus becomes a stochastic process (see Fig.~\ref{fig:domains}) that can be simulated in a double Monte Carlo fashion: first, for a given, randomly sampled  initial velocity distribution, the initial topological charges are distributed randomly. Then the probabilities of transmissions and reflections are computed for each collision, which are then  used to generate 
 sets of topological  charge configurations with appropriate Monte Carlo weights.

In the extreme limits of perfect reflection at each collision (original semiclassical approach), and perfect transmission (noninteracting particles) realized at the free fermion point, $\gamma=1/\sqrt{2},$ charge averaging along the trajectories  is not necessary, since instantaneous charge
 configurations are uniquely determined by the initial charge and velocity configuration. In this limit, analytical results can be obtained~\cite{Damle2005,Rapp2006}.

\section{Dynamical correlation function}
\label{sec:corr}

We are interested in the finite temperature dynamical two-point function of general vertex operators \erf{eq:corrfn}.
In the semiclassical approach, this is replaced by the average\footnote{The difference between Eqs. \erf{eq:corrfn} and \erf{eq:C} is expected to cause, at most, a power law of $t$ deviation in the asymptotic long-time limit.} 
\be
\label{eq:C}
C_\eta(x,t)=\big\langle e^{i\eta \left(\Phi(x,t)-\Phi(0,0)\right)}\big\rangle_T\,.
\ee
In the semiclassical approach, we compute the expectation value \erf{eq:C}
by averaging over  semiclassical histories. In a given history, the quantity within the average 
is measured by noticing that within space-time domains bordered by kink trajectories, the field sits in one of the classical vacua, i.e., it takes constant values, $\Phi(x,t)=(2\pi/\gamma)Q,$ where the integer charge of the domain, $Q\in\mathbb{Z},$ depends on the initial charge
configuration as well as on the zig-zag trajectories of the charges. Consequently, 
\be
\label{eq:C2}
C_\eta(x,t)=\big\langle e^{i\eta \left(\Phi(x,t)-\Phi(0,0)\right)}\big\rangle_T =\big\langle e^{i\bar\eta\Delta Q(x,t)}\big\rangle_T\,,
\ee
where
\be
\bar\eta = 2\pi\eta/\gamma\,,
\ee
and $\Delta Q(x,t)=Q(x,t)-Q(0,0)$ is the charge difference of the domains containing the points $(0,0)$ and $(x,t).$

\subsection{Analytical results: diffusive and ballistic limits}
\label{sec:analytic}

The average \erf{eq:C} can be evaluated numerically in a Monte Carlo simulation. 
Before discussing our numerical results, however, we focus on  
 the purely reflective and transmissive limits, where we can make 
analytic progress, and predict the asymptotic long time and large distance
behavior of the correlation function.

In the first steps,  we follow the derivation of Ref. \cite{Damle2005}, and  consider the segment $[(0,0),(x,t)]$
connecting the two operator insertions in the space-time diagram of Fig. \ref{fig:domains},
and crossing some domains delineated by kink trajectories. The charge difference $\Delta Q(x,t)$ can be determined based 
on the number and charges of the kink lines crossing this segment. Whenever the segment $(0,0)\to(x,t)$
crosses a soliton line that comes  from the right, we enter a new domain of topological charge $+1$ with respect to the one we were in before the crossing. For soliton lines from the left, the charge difference is $-1.$ For antisolitons, there is an extra minus sign in both cases.

To intersect the segment $[(0,0),(x,t)]$,  a world line from the right  must have been within the spatial interval $[0,x-vt]$ at $t=0$.
 Similarly, a world line crossing from the left must have started in the interval $[x-vt,0].$ 
The average number of world lines crossing from the right and from the left is given by
\begin{align}
\pr (x,t)&\equiv\langle N_\text{r}(x,t)\rangle= \rho \int_{-c}^c \ud v f(v) \Theta(x-vt)(x-vt)\,,
\nonumber
\\
\pl(x,t) &\equiv\langle N_\text{r}(x,t)\rangle= \rho  \int_{-c}^c \ud v f(v) \Theta(vt-x)(vt-x)\,.
\nonumber
\end{align}
Since  world lines are statistically independent of each other and are distributed uniformly in space, the left and right intersections are independent Poisson processes, so the probability of $\kl$ left and $\kr$ right intersections is
\be
\label{eq:Pklkr}
P(\kl,\kr) = \frac1{\kl!\kr!}\pl^{\kl}\pr^{\kr} e^{-(\pl+\pr)}\,.
\ee
The correlation function can then be expressed as
\be
\label{eq:Cgen}
C_\eta(x,t) = \sum_{\kl=0,\kr=0}^\infty P(\kl,\kr) \vev{e^{i\bar\eta\Delta Q(\kl,\kr)}}_\text{ch}\,.
\ee
Here $\Delta Q(\kl,\kr)$ is the difference of the topological charges of the domains where the points $(0,0)$ and $(x,t)$ lie and which are separated by $\kl$ and $\kr$ left and right crossing kinks, and the average $\vev{\dots}_\text{ch}$ is over the initial topological charges at fixed $\kl$ and $\kr.$
In the purely reflective or transmissive cases, $\Delta Q(\kl,\kr)$ is completely 
 determined by the initial charge configuration, 
and the charge averaging can be carried out analytically. In the general case, however, 
$\Delta Q$ is a stochastic variable even for a fixed initial charge configuration, and 
averaging implies a Monte Carlo simulation.

\subsubsection{Perfect transmission or reflection}

The average in Eq. \erf{eq:C2} must be taken over the initial positions and velocities of the kinks, and over the charge history.

For the \emph{perfectly transmitting} case, the topological charges crossing the segment $(0,0)\to(x,t)$ are completely uncorrelated. 
Each crossing kink is either a soliton or antisoliton with probability $1/2,$ and either comes from the left or from the right with probability $1/2,$ so 
%
 $\vev{e^{i\bar\eta\Delta Q(\kl,\kr)}}_\text{ch}^\text{(tr)}
 = \left(\frac12e^{i\bar\eta}+\frac12e^{-i\bar\eta}\right)^{(\kl+\kr)} = (\cos\bar\eta)^{(\kl+\kr)}.$
%
The distribution of the sum of two Poissonian variables is again Poissonian, so
\be
\label{eq:Ctr1}
C_\eta^\text{(tr)}(x,t) = \sum_{n=0}^\infty P^\text{(tr)}(n) (\cos\bar\eta)^n
\ee
with $P^\text{(tr)}(n=\kr+\kl)=\frac{(\pr+\pl)^n}{n!}e^{-(\pr+\pl)}$ which gives \cite{Damle2005}
%
\be
\label{Ctr}
C_\eta^\text{(tr)}(x,t) = e^{-(1-\cos\bar\eta)(\pl+\pr)} = e^{-2\rho\sin^2(\frac{\bar\eta}2)\!\int_{-c}^{c}\! \ud v f(v) |x-vt| }\,.
\ee
%
In particular, both the autocorrelation function $C_\eta(0,t)$ and the static correlation function $C_\eta(x,0)$ 
decay {\em exponentially} in the transmissive case:
\bes
\begin{align}
C_\eta^\text{(tr)}(0,t) &= e^{-2\sin^2(\frac{\bar\eta}2)t/\tau} \,,\label{eq:autoCtr}\\
C_\eta^\text{(tr)}(x,0) &= e^{-2\sin^2(\frac{\bar\eta}2)\rho \,x} \,,
\end{align}
\esu
where $\rho$ is the kink density and $\tau$ is the characteristic collision time
defined in Eq. \erf{eq:tau}.

In the other extreme case of \emph{perfectly reflective} collisions, the calculation is more complicated as the trajectories of the charges become correlated. 
The key observation is that the spatial sequence of charges of the domains from left to right at any given time is invariant under the time evolution \cite{Damle2005,Rapp2006}. As a consequence, the charge difference only depends on the ``distance'' 
of the domains of $(0,0)$ and $(x,t)$ in the sequence of domains,  
\be
\label{eq:s}
s=\kr-\kl\;,
\ee
which also determines $\Delta Q(\kl,\kr)=\Delta Q(\kr-\kl).$ 

Inserting a factor of 1 in the form
%
$\sum_{s=-\infty}^\infty \delta_{\kr-\kl,s} = \sum_{s=-\infty}^\infty\int_0^{2\pi}\frac{\ud\phi}{2\pi} e^{i\phi(\kr-\kl-s)}$
in Eq.~\eqref{eq:Cgen},
%
we can perform the sum over $\kl$ and $\kr$:
\begin{multline}
\label{eq:corr}
C^\text{(r)}_\eta(x,t) =  \sum_{s=-\infty}^\infty\int_0^{2\pi} \frac{\ud\phi}{2\pi} e^{-i\phi s}\vev{e^{i\bar\eta\Delta Q(s)}}_\text{ch}\\
 e^{-\pl(x,t)(1-e^{-i\phi})-\pr(x,t)(1-e^{i\phi})}\,.
\end{multline}
The integral can be evaluated giving
\be
\label{eq:corr2}
C^\text{(r)}_\eta(x,t) = \sum_{s=-\infty}^\infty \mathcal{P}(s) \vev{e^{i\bar\eta\Delta Q(s)}}_\text{ch}
\ee
with
\be
\label{eq:Ps}
\mathcal{P}(s=\kl-\kr) = e^{-(\pl+\pr)}(\pr/\pl)^{s/2}I_{|s|}(2\sqrt{\pr\pl})\,,
\ee
where $I_n(x)$ denotes the modified Bessel function. As expected, we recovered the so-called Skellam distribution obeyed by the difference of two Poisson distributed random variables.

In this reflective limit, due to the conservation of the domain charges, the charge difference $\Delta Q(s)$ does not change in time, and can be evaluated at $t=0.$ The charges are independent and take $\pm1$ with probability $1/2,$ so
 %
$\big\langle e^{i\bar\eta\Delta Q(s)} \big\rangle =  (\cos\bar\eta)^{|s|}$ which leads to
\be
C_\eta^\text{(r)}(x,t) = \! \sum_{s=-\infty}^\infty \!e^{-(\pl+\pr)}\!\left(\frac\pr\pl\right)^{\!\frac{s}2}\!\!I_{|s|}(2\sqrt{\pr\pl}) (\cos\bar\eta)^{|s|}\,.
\ee
We can obtain another representation by exchanging the integral and the sum in Eq. \erf{eq:corr} and evaluating the latter, yielding
\begin{multline}
\label{eq:Cr}
C_\eta^\text{(r)}(x,t) = \int_0^{2\pi} \frac{\ud\phi}{2\pi} 
\frac{1-\cos^2\bar\eta}{1-2\cos\bar\eta\cos(\phi)+\cos^2\bar\eta}\\
e^{-2\sin^2(\phi/2)(\pr+\pl)+i\sin(\phi)(\pr-\pl)}\,.
\end{multline}
In the case of the autocorrelation function we have $\pl(0,t) = \pr(0,t)=t/(2\tau),$ and the last exponential reduces to $e^{-2\sin^2(\phi/2)t/\tau}.$ A similar, but somewhat different expression was obtained in Ref. \cite{Altshuler2006} using form factor techniques.

\begin{figure}[t]
\includegraphics[width=0.4\textwidth]{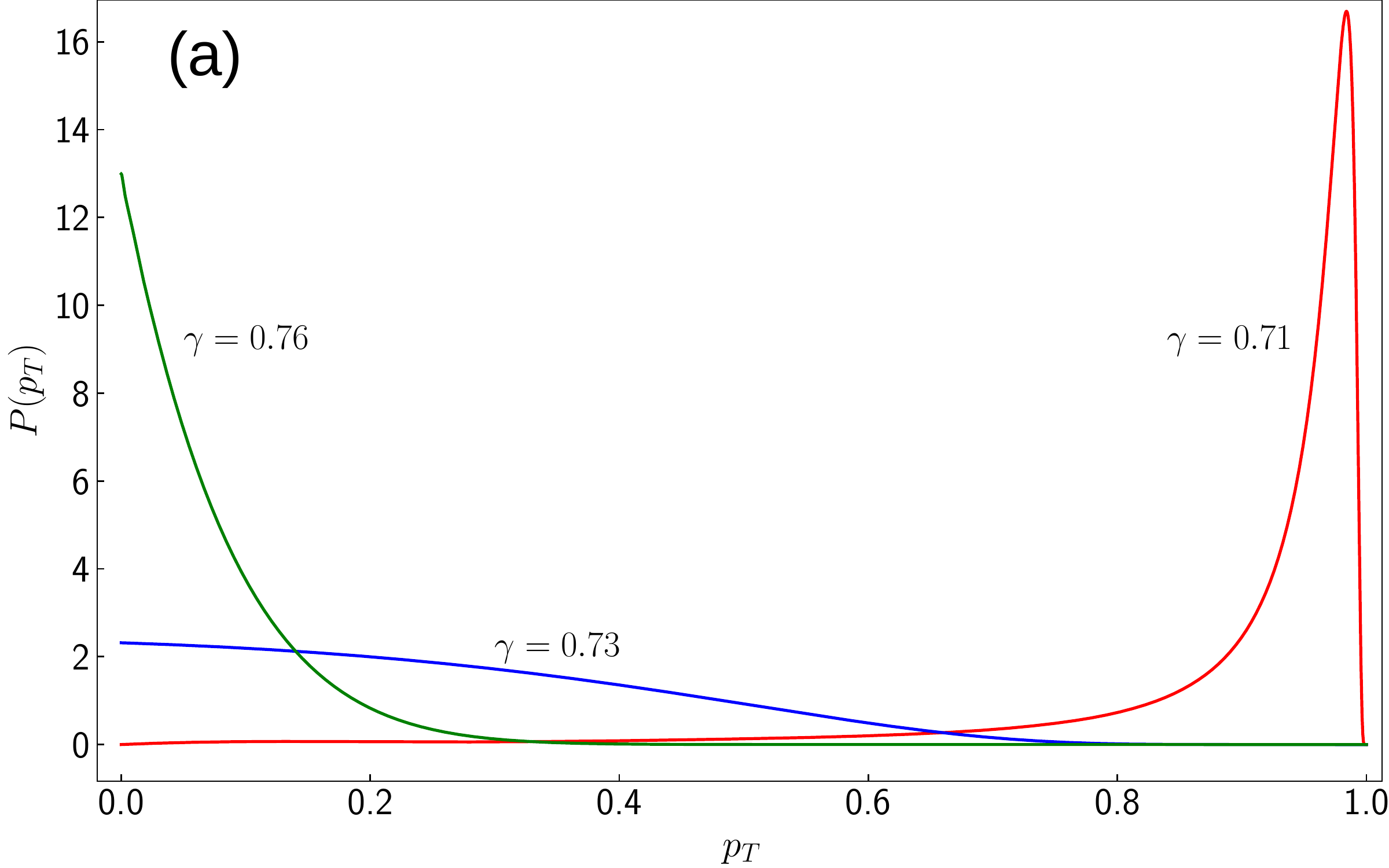}\\
\includegraphics[width=0.4\textwidth]{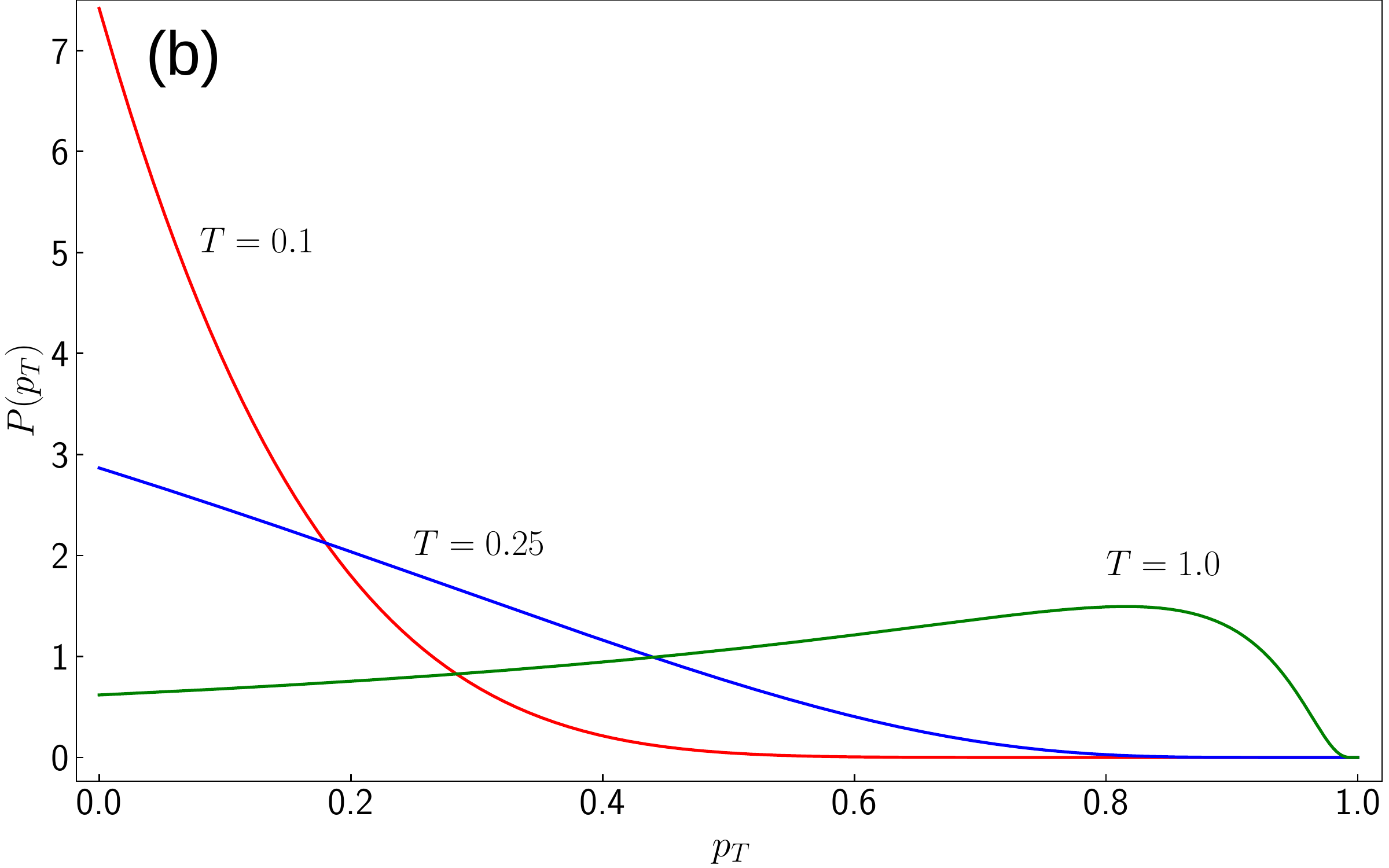}
\caption{Probability distribution $P(\pt)$ of the transmission probability $\pt.$ (a) $P(\pt)$ for three different values of the coupling $\gamma$ at temperature $T=0.02Mc^2.$ (b) $P(\pt)$ for three different temperatures for coupling $\gamma=0.8.$ \label{fig:pT}}
\end{figure}

Outside the light cone, $x>c\,t,$ left crossings are impossible: $\pl=0,$ $\pr=\rho x,$ and $\kl=0.$\footnote{For $x<-c\,t,$ analogous statements hold with left and right quantities interchanged.} 
This implies that $n=|s|$ and $P(s)=P^\text{(tr)}(|s|)$, so $C_\eta^\text{(tr)}(|x|>c\,t)=C_\eta^\text{(r)}(|x|>c\,t)=e^{-2\sin^2(\bar\eta/2)\rho|x|}$,
independent of time, a property reflecting  the light cone effect. 
It is also easy to check that for short times $t\ll\tau,$ before the collisions can make their effect, $C_\eta^\text{(tr)}(0,t)\approx C_\eta^\text{(r)}(0,t)\approx 1-2\sin^2(\bar\eta/2) t/\tau.$

\begin{figure*}[t]
\centering
\includegraphics[width=0.49\textwidth]{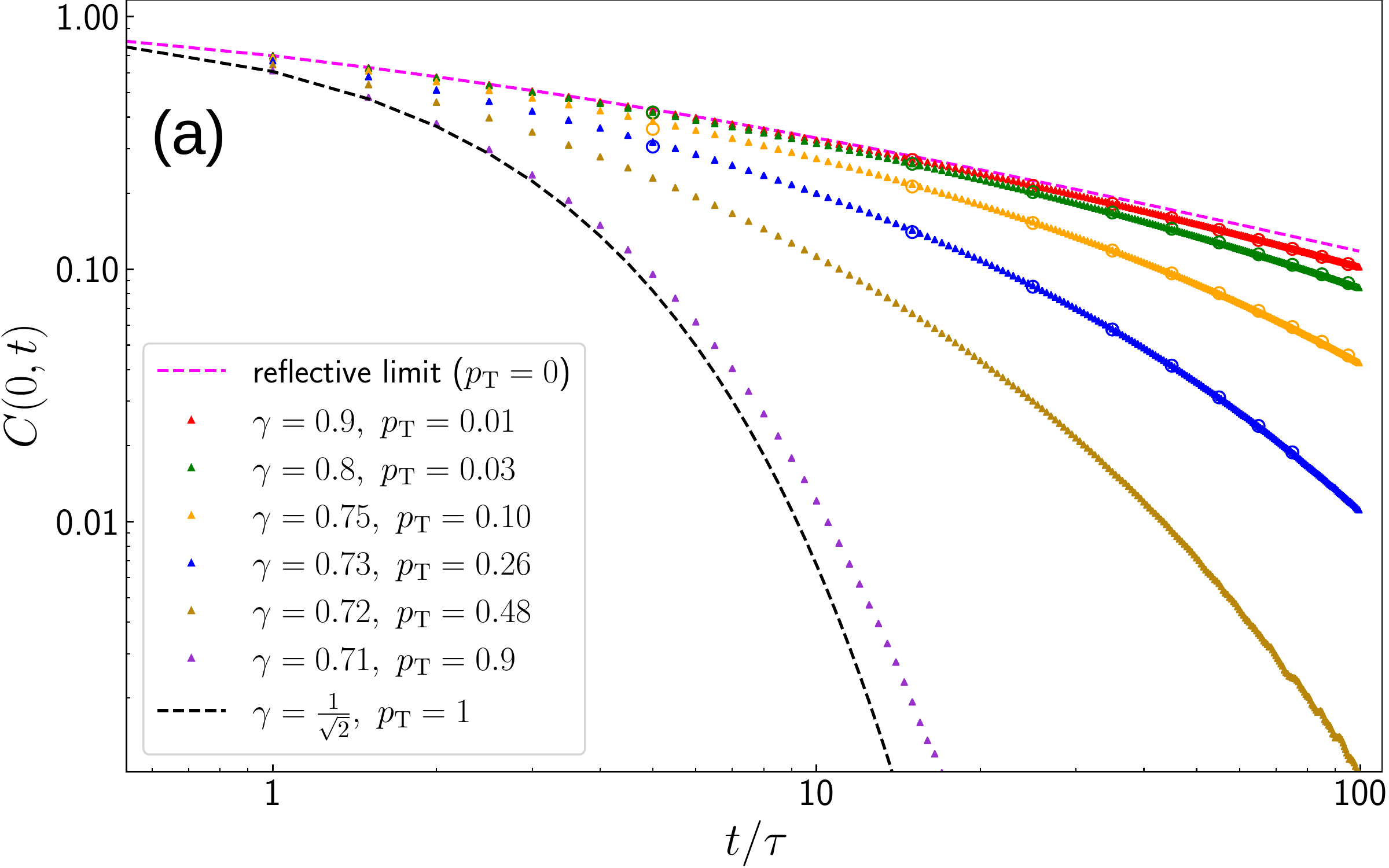}
\hfill
\includegraphics[width=0.49\textwidth]{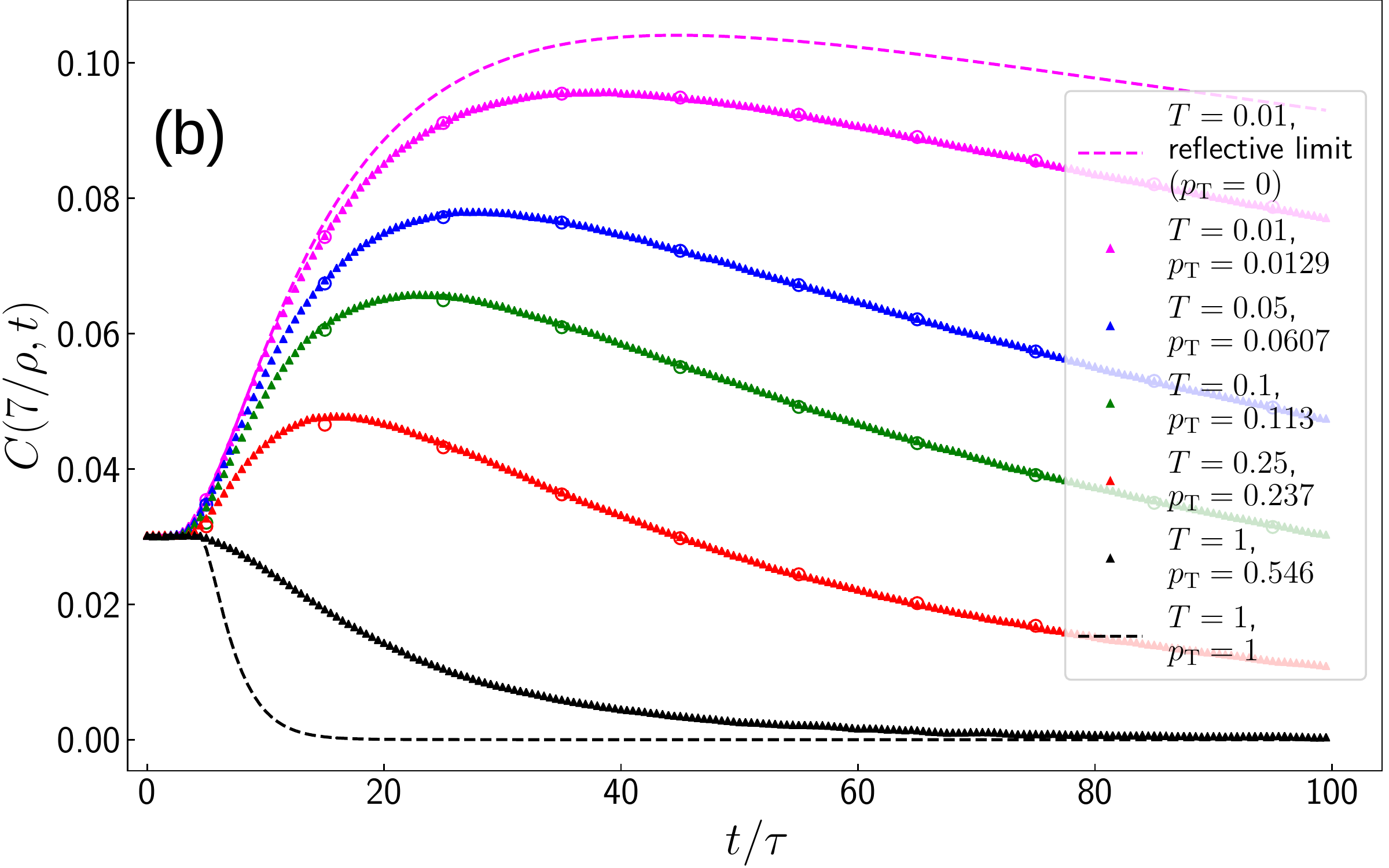}
\caption{Correlation functions $C_\eta(x,t)$ for $\bar\eta=\pi/3$. (a) Autocorrelation function $C_\eta(0,t)$ on a log-log scale as a function of dimensionless time $t/\tau$ at temperature $T=0.02\,Mc^2$ for $\gamma=0.71,0.72,0.73,0.75,0.8,0.9$ (from bottom to top). 
(b) Correlation function $C_\eta(x=7/\rho,t)$ at $\gamma=0.8$ 
at different temperatures $T/(Mc^2)=0.01,0.05,0.1,0.25,1$ (from top to bottom).
In both panels, the dots correspond to the semiclassical Monte Carlo simulation results, the empty circles represent the SEP results, 
while the perfect transmission (bottom) and perfect reflection (top) limits are shown in dashed lines. \label{fig:corr} }
\end{figure*}

However, the large time behavior inside the light cone, $|x|<c t,$ is very different in the two cases. We can obtain the asymptotics of Eq. \erf{eq:Cr} using the saddle point approximation. We note that $\pr-\pl = \rho x$ exactly, while $\pr+\pl\approx t/\tau +M/(2T)x^2/(t\tau)$ for $x\ll t.$ So for large $t\gg\tau,x,$ the exponent is dominated by the real term, and the saddle point is at $\phi=0.$ This leads to the {\em diffusive} form \cite{Damle2005}
\be
\label{eq:Crasym}
C_\eta^\text{(r)}(x,t)
\approx\frac{ \cot^2\left(\frac{\bar\eta}2\right) e^{-\frac{x^2}{4Dt}}}{\rho\sqrt{4\pi Dt}}
\ee
with the diffusion constant
\be
D=\frac1{2\rho^2\tau} \approx
\frac{e^{Mc^2/T}}M\,,
\ee
where the approximation is valid for $T\ll Mc^2.$
For large $t$ and $x^2/t$ fixed, $x\ll t$ holds, consistently with our assumption in the saddle point calculation. Note that 
the autocorrelation function decays algebraically in the reflective case, as $\sim1/\sqrt t$.

\subsection{Numerical results}

After discussing the two extreme, analytically tractable  cases, we now turn to the numerical evaluation of Eq. \erf{eq:C} within the semiclassical framework. In our simulations, we used units $c=k_\text{B}=\hbar=1$ and $M=1,$ and  measured dimensionful quantities in appropriate powers of the soliton mass $M.$ 

We first set the initial state by placing $N$ atoms randomly in a box of length $L,$ and 
assigning $q=\pm1$ 
charges to them with equal probability. Their initial velocities are
then drawn from the classical, relativistic thermal distribution, \erf{fv}. This fixes the straight kink trajectories and the positions and times of the collisions up to some maximal time (e.g., the time separation of the correlation function). The history of the charges is then determined by going 
through the collisions one by one in chronological order. For each collision, the transmission and reflection probabilities, $|S_\text{T}(\theta_1-\theta_2)|^2$ and $|S_\text{R}(\theta_1-\theta_2)|^2,$ are computed from the rapidities $\theta_1,\theta_2$ of the incoming particles using Eq.~\erf{eq:ST}, and then one of the two possible outcomes is selected in a Monte Carlo step. Finally, the charge difference $\Delta Q$ appearing in Eq.~\erf{eq:C2} is  computed from the kink charges. Averaging is then performed by repeating this procedure  around $25,000-95,000$  times.

To  get rid of finite size effects, we have chosen a system size $L\gg ct$.
In this way, the measurements can be done far away from the edges (a distance $ct$ is sufficient) and the autocorrelation function 
can be measured more efficiently by performing a spatial average.

First we show in Fig. \ref{fig:pT} the distribution of the transmission probability over the collisions.  
At the free fermion point, $\gamma=1/\sqrt2\approx0.707,$ we have perfect transmission, and the distribution is a Dirac-delta, $P(\pt)=\delta(\pt-1).$ However, the distribution broadens and its average  decreases rapidly upon increasing $\gamma,$ and for even higher values of $\gamma$ its maximum is at $\pt=0.$ We can thus interpolate between the two extreme limits, the fully transmissive and fully reflective cases, by changing the sG coupling $\gamma.$ 
Conversely, fixing $\gamma$ and varying the temperature $T,$ the distribution changes from having a maximum at $\pt=0$ (low $T$) to being peaked near $\pt=1$ (high $T$).

Fig. \ref{fig:corr}  shows the results of the semiclassical ``molecular dynamics'' simulation for the autocorrelation function $C_\eta(x=0,t)$ vs. $t/\tau$ for $\bar\eta=\pi/3$ ($\eta/\gamma=1/6$) at a temperature $T=0.02\,Mc^2$ (small triangles). 
Different colors correspond to different values of $\gamma.$ As $\gamma$ is increased, 
the correlation function crosses over from a purely transmissive exponential behavior 
 to the purely reflective $\sim1/\sqrt{t}$ result (both  shown as dashed lines).
Between these two curves, the asymptotic time dependence of the semiclassical results is not a pure power law, and it can also be checked that it is not an exponential decay. We shall derive an approximate formula for this time dependence in the next subsection.

In the right panel of Fig. \ref{fig:corr},  we plot our numerical results for the correlator at finite spatial separation $x=7/\rho$
for $\bar\eta=\pi/3.$ Here we varied the temperature at fixed coupling $\gamma=0.8$ in order to navigate between the 
the purely reflective (low $T$) and transmissive (high $T$) cases which are again plotted in dashed lines. As the temperature is lowered, a peak develops in the time dependence of the correlation function,  similar to the diffusive result \erf{eq:Crasym}. The generic long time behavior is apparently
qualitatively different from both analytic cases.

\subsection{Mapping to the simple exclusion process}

At small enough temperatures, the momenta of the kinks are small, and their collision is predominantly reflective
for any $\gamma\ne1/\sqrt{2}$. However, transmissive processes are also present, and while
the original semiclassical approach  assuming perfect reflection may be a good approximation for some time, 
it is expected to break down for sufficiently long times. 
 
The time scale at which the original semiclassical approach breaks down can be estimated as follows.
 At low temperatures, transmissive collisions  have a small probability, $\pt\sim \bar v^2\sim T/M.$
In a time period $t,$ a given kink suffers $t/\tau_\text{c}$ collisions on average, where $\tau_\text{c}$ is the 
collision time of the order of $\tau$. This means that after time $t^*\sim \tau/\pt \sim M/T^2e^{Mc^2/T},$ the 
charges of the kinks will change, and their sequential order, the origin of diffusive scaling, will not be preserved.

In the generic case, an analytic calculation similar to the ones in Sec. \ref{sec:analytic} is not possible,
 because  trajectory and charge averages do not factorize, in general.
The fate of the charges and the space-time properties of the trajectories get correlated, because the 
outcome of each collision is not fixed but depends on the incoming velocities (slopes of the trajectories).

However, we can make some analytic progress by focusing on the limit of small transmission
 probability, and by  replacing the velocity dependent transmission probability by its mean value, $\pt$. 
We  assume furthermore that -- since transmissive collisions happen very rarely in our limit -- transmissions of neighboring charges happen randomly in the space-time history. The average charge difference of the domains of the two operators does not depend exclusively on the  intersection numbers anymore, but it becomes effectively time dependent due to  rare transmissive collisions that change the charges of the domains. With these assumptions, the charge and trajectory averages  factorize, and Eqs. \erf{eq:corr} and \erf{eq:corr2} can be used after replacing $\Delta Q(s)$ by the effectively time-dependent $\Delta Q(s;t).$

The dynamics of the soliton charges in this simplified model can be mapped to the Simple Symmetric Exclusion Process (SEP), a classical Markovian stochastic process describing hardcore particles hopping on a lattice. The mapping is illustrated in Fig. \ref{fig:domains}. The sites of the 1D lattice of the SEP correspond to the kinks. A site is occupied if the kink is a soliton and empty if it is an antisoliton. 
In the SEP, particles attempt a jump at a constant rate $p\propto \pt$ with equal probability to the left and to the right,
 but can only jump if the target site is empty. This is mapped to the collision of kinks: a particle jump corresponds to swapping 
 a soliton and an antisoliton with probability $\pt.$  The distance that a SEP particle travels 
 corresponds to $s$ given by Eq. \erf{eq:s}, the distance in terms of the domains.

Finally, we have to match the time scales of the two systems. The rate of
 jumps in the SEP is related to the frequency of transmissive soliton-antisoliton collisions 
 which depends on the mean collision time as well as on the transmission probabilities. 
 Away from the perfectly transmissive and reflective limits, the distribution of transmission probabilities 
 is typically very broad (c.f. Fig. \ref{fig:pT}) and there is no unambiguous way to 
 capture it by a single number (e.g., the mean). For this reason, we leave the proportionality factor between the physical and the dimensionless
 SEP time as a free parameter $\alpha\propto \pt,$
\be
\label{alphadef}
\hat t = \alpha \frac{t}\tau\,.
\ee

\begin{figure}[t]
\centering
\includegraphics[width=0.49\textwidth]{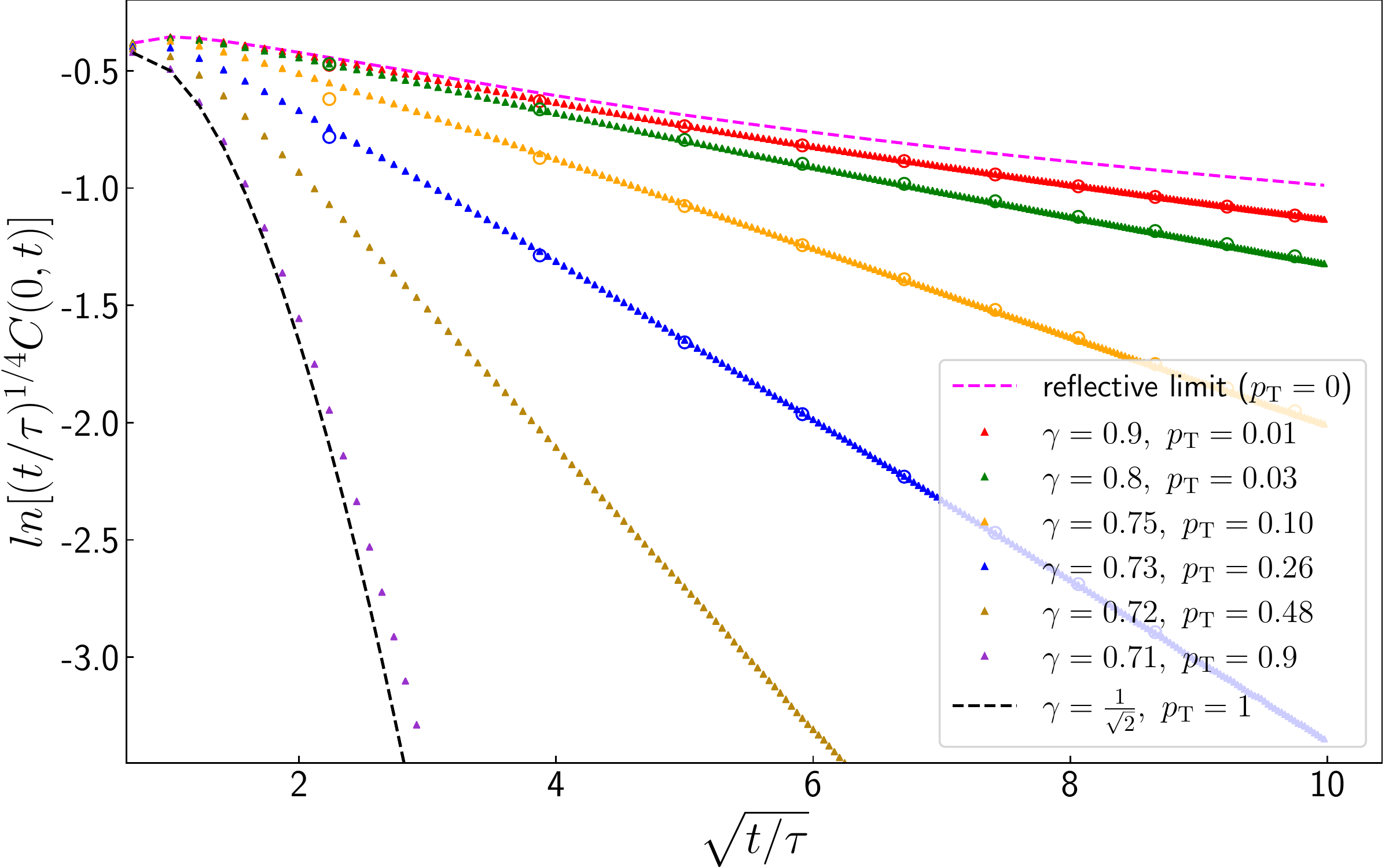}
\caption{
Rescaled autocorrelation function $\ln[(t/\tau)^{1/4}C(0,t)]$ against $\sqrt{t/\tau}.$ 
The data and the conventions are the same as in Fig. \ref{fig:corr}(a). \label{fig:rescaled}}
\end{figure}

The mapping to the SEP allows us to use some of the exact results in this model. In particular,
 an analytic result has been derived in Ref. \cite{Imamura2021} for the cumulant generating 
 function of a quantity closely related to the transferred charge $\Delta Q_s(\hat t)$ in terms 
 of a Fredholm determinant (for the explicit expressions, see Appendix \ref{app:SEP}). This result allows 
 us to compute $\big\langle e^{i\bar\eta\Delta Q_s(\hat t)}\big\rangle$ numerically for any $s$ and $\hat t$, which 
 then can be used for the numerical evaluation of the correlator in Eq. \erf{eq:corr2}.

The large time asymptotics has also been derived in Ref.~\onlinecite{Imamura2021}. Translated into 
our language, in the limit $\hat t\to\infty,$ $s\to\infty,$ $\xi\equiv-s/\sqrt{4\hat t}$ fixed, we have
\be
\label{eq:SEPas}
\vevn{e^{i\bar\eta\, \Delta Q_s(\hat t)}}_\text{ch} \sim 
e^{-\sqrt{\hat t}\, \w(\xi)}\,,
\ee
where
$ \w(\xi) = \sum_{n=1}^\infty \frac{\sin^{2n}(\bar\eta)}{n^{3/2}} 
A(\sqrt{n}\xi)$
with 
$ A(y) = \frac{e^{-y^2}}{\sqrt\pi} + y\, \mathrm{erf}(y). $

It is the Fourier series of the expectation value \erf{eq:SEPas}
 that appears in Eq. \erf{eq:corr}. 
In the large $t$ limit, the final integral over $\phi$ will be evaluated again in the saddle point approximation, 
where the saddle point remains at $\phi=0$ since the exponent of \erf{eq:SEPas} is $\sim\sqrt{t}$ as opposed to $\sim t$. 
Setting $\phi=0$ and approximating the sum over $s$ by an integral over $\xi,$ we obtain
\be
\sum_{s=-\infty}^\infty\vevn{e^{i\bar\eta\, \Delta Q_s(\hat t)}}_\text{ch} \approx
-2\sqrt{\hat t} \int \ud \xi e^{-\sqrt{\hat t}\,\omega(\xi)}\approx a\,\hat t^{1/4}e^{-b\sqrt{\hat t}}\,,
\ee
where we evaluated the integral in the saddle point approximation, yielding  
$a=-2\sqrt{{2\pi}/{\w''(0)}}=\frac{-2\pi^{3/4}}{\sqrt{\mathrm{Li}_{1/2} \left[\sin^2(\bar\eta) \right]}} $
and 
$ b=\w(0)=\frac1{\sqrt\pi} \,\mathrm{Li}_{3/2} \left[\sin^2(\bar\eta) \right] $
in terms of the polylogarithm function $\mathrm{Li}_n(x).$
Plugging this result in Eq. \erf{eq:corr} and performing the $\phi$-integral 
in the saddle point approximation, we obtain
\be
\label{eq:SEPasym}
C_\eta(x,t)\approx  \bar a \,t^{1/4}\; e^{-\bar b \sqrt{t}} \;\frac{e^{-x^2/(4Dt)}}{\rho\sqrt{4\pi D t}}
\ee
with 
$\bar a=a\,\alpha^{1/4}$
and
$\bar b= b\,\alpha^{1/2}.$ 
\footnote{This result can also be obtained by using a Gaussian approximation for the distribution $P(s)$ in Eq. \erf{eq:corr2} (obtained, e.g., by performing the $\phi$-integral in Eq. \erf{eq:corr} in the saddle point approximation), replacing the sum over $s$ by an integral over $\xi,$ and evaluating the integral in the saddle point approximation using that $\pr+\pl\sim t \sim x^2\sim (\pr-\pl)^2.$}

Eq. \erf{eq:SEPasym} is one of the central results of our paper. It shows that the $\sim1/\sqrt t$ diffusive behavior
 found for perfect reflections is not universal, but gets modified for large enough times in the presence of 
 an arbitrary small transmission rate. The more generic result is a {\em stretched exponential}, 
 $\sim e^{-\bar b \sqrt t},$ with a slowly decaying $\sim t^{-1/4}$ prefactor.

To verify  the asymptotic time dependence predicted by Eq.~\erf{eq:SEPasym}, 
we display the rescaled correlation function, $\ln (t/\tau)^{1/4}C_\eta (0,t)$ vs. $\sqrt{t/\tau}$ in Fig. \ref{fig:rescaled}. 
As expected, the SEP approximation  
 (empty circles) becomes very accurate as the reflective limit is approached, and 
   gives a good approximation 
for   $\gamma\gtrsim 0.75$, where $\pt \lesssim 0.1$.

Moreover, the SEP result  appears to capture the asymptotic stretched exponential behavior correctly for any $\gamma\ne1/\sqrt{2}$.  
The plotted curves indeed seem to be straight lines starting from an early time even for larger transmission rates, 
indicating that the time dependence of both the exponent and the prefactor in Eq. \erf{eq:SEPasym} (for $x=0$) hold beyond 
the range of validity of Eq. \erf{eq:SEPasym}. 
This suggests that the stretched exponential decay is the universal behavior instead of the special 
results found in the extreme limits of perfect transmission or reflection.

\section{Full counting statistics}
\label{sec:PQ}

As Eq. \erf{eq:C2} shows, the autocorrelation function
is the moment generating function of the distribution of the net charge transferred through the origin,
\be
Q_0(t)\equiv \Delta Q(0,t)  = Q(0,t) - Q(0,0)\,.
\ee
The distribution of $Q_0(t)$ is just given by
\be
\label{eq:Pgen}
P(Q_0;t) = \sum_{\kl=0,\kr=0}^\infty P(\kl,\kr) P_{\kr,\kl}(Q_0;t)\,,
\ee
where $P(\kl,\kr)$  is given in Eq. \erf{eq:Pklkr}, and $P_{\kr,\kl}(Q_0;t)$ is the (conditional) 
probability of $Q_0$ for $\kr$ right and $\kl$ left crossings.

\subsection{Charge distributions in the reflective and transmissive limits}

\begin{figure*}[t]
\centering
\includegraphics[width=0.49\textwidth]{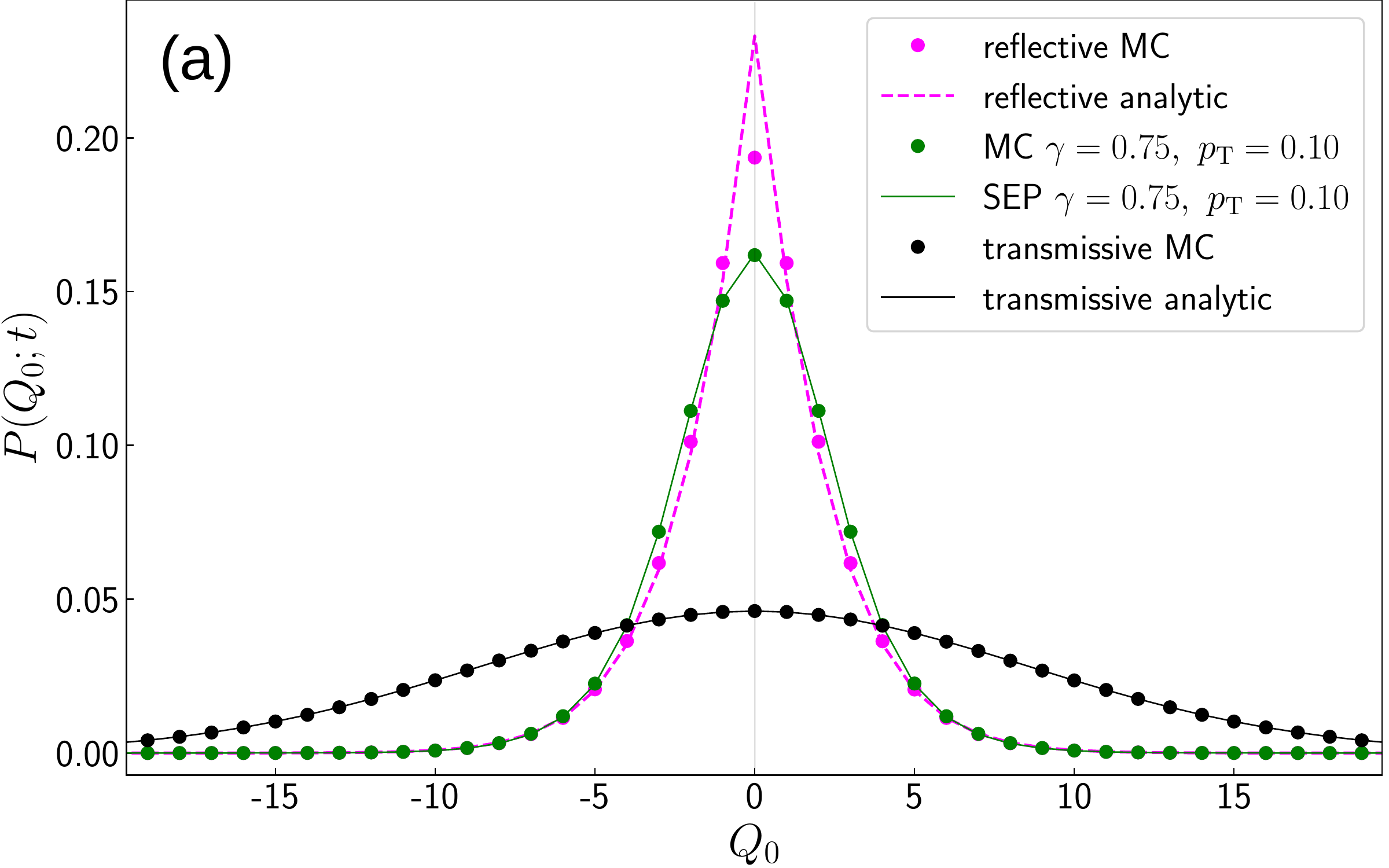}
\hfill
\includegraphics[width=0.49\textwidth]{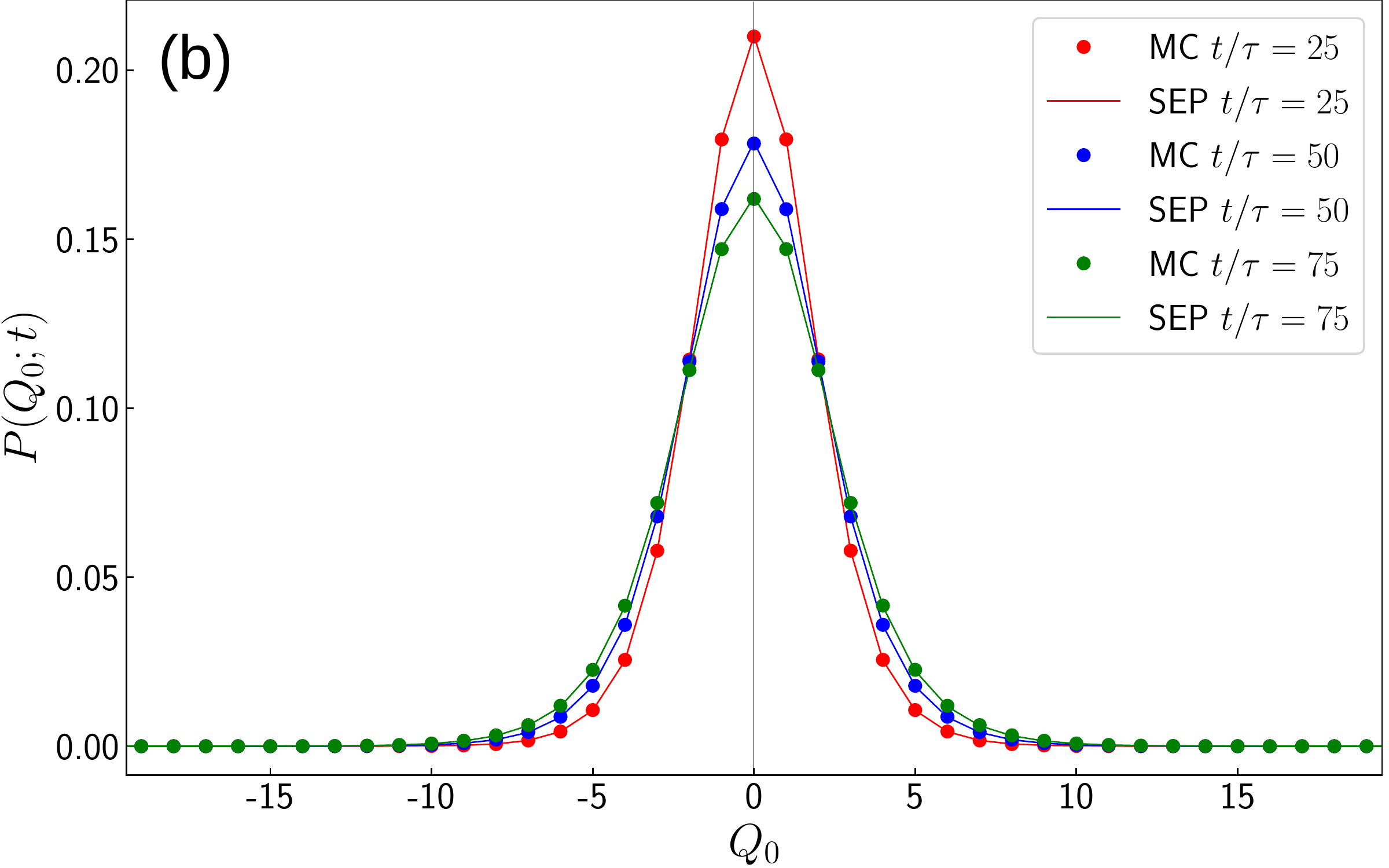}
\caption{Full counting stastistics of the topological charge transfer. (a) Full distribution functions of the transferred charge at temperature $T=0.02Mc^2$ at time $t/\tau=75$ in the purely transmissive  (black) and purely reflective (magenta) limits, and in the gerneric case with finite transmission at $\gamma=0.73$ ($\pt\approx0.1$). Symbols represent the semiclassical simulation results. The black line shows the exact transmissive result \erf{eq:PQtr}, the green line represents the SEP approximation, and the magenta line displays the approximate result \erf{eq:PQras}.
(b) Full distribution functions of the transferred charge at temperature $T=0.02Mc^2$, $\gamma=0.75,$ and times $t/\tau=25,50,75.$ The simulation results are shown in dots, the lines correspond to the SEP approximation.  \label{fig:PQ} }
\end{figure*}

In the perfectly transmitting case, the conditional probability $P_{\kr,\kl}(Q_0;t)$ is, of course, the binomial distribution,
\be
\label{eq:binom}
P^\text{(tr)}_{\kr,\kl}(Q_0;t) 
=  \frac1{2^{\kr+\kl}}\binom{\kr+\kl}{\frac{Q_0+\kr+\kl}2},
\ee
if $Q_0+\kr+\kl$ is even and nonnegative, and 0 otherwise. Substituting this in Eq. \erf{eq:Pgen} we obtain
\be
\label{eq:PQtr}
P^\text{(tr)}(Q_0;t) = e^{-t/\tau} I_{Q_0}(t/\tau)\,,
\ee
where we used that for $x=0$ the left and right Poisson parameters are equal,
$\pr(0,t)=\pl(0,t)=t/(2\tau)$.\footnote{This is again a Skellam distribution, because $Q_0$ is the difference of Poisson distributed left and right crossings of positive and negative charges. It can also be obtained by taking the Fourier transform of Eq. \erf{eq:autoCtr}. } The distribution is plotted in Fig. \ref{fig:PQ}(a) for  $t/\tau=75.$ For  $t\gg\tau,$ the distribution approaches a Gaussian of mean deviation $\sqrt{t/\tau}$,
and $P^\text{(tr)}(Q_0;t)\sim (2\pi\, t/\tau)^{-1/2}$ for any fixed $Q_0$.

For perfect reflections, the conditional probability $P^\text{(r)}_{\kr,\kl}(Q_0;t)=P^\text{(r)}_s(Q_0)$ is also the
 binomial distribution, Eq.~\erf{eq:binom}, but with the replacement  $\kr+\kl\to|s|=|\kr-\kl|$.
For $\pr=\pl=t/(2\tau)$ Eq.~\eqref{eq:Ps}
reduces to 
\be
{\cal P}(s) = e^{-t/\tau} I_{|s|}(t/\tau)\,,
\ee
and Eq.~\eqref{eq:Pgen} can be expressed as
\be
\label{eq:PQr}
P^\text{(r)}(Q_0;t) = \sum_{s=-\infty}^\infty  e^{-t/\tau} I_{|s|}(t/\tau)  \,\frac1{2^{|s|}}\binom{|s|}{\frac{Q_0+|s|}2}\,.
\ee

We are interested in the large time asymptotic behavior of the distribution $P^\text{(r)}(Q_0;t)$. 
Unfortunately, it cannot be determined by taking the Fourier transform of the asymptotic results in Eqs. \erf{eq:Crasym} 
and \erf{eq:SEPasym}
because the integrals do not converge. 
We use the representation \erf{eq:Cr} instead and exploit that the Fourier transform of the integrand  can be written in terms of 
hypergeometric functions (c.f. Appendix \ref{app:FCSrep}).\footnote{We derived this by evaluating the sum $\sum_s e^{-i\phi s} P^\text{(r)}_s(Q_0)$ coming from Eq. \erf{eq:corr}.}
Expanding them around $\phi=0$ and evaluating the $\phi$-integral in the saddle point approximation we obtain for the large time behavior
for $|Q_0|\ll (t/\tau)^{1/4}$
\begin{multline}
\label{eq:PQrepas}
P(Q_0;t)^\text{(r)} \approx 
\frac{\Gamma(\frac14)}{2\pi}\left(\frac{2}{t/\tau}\right)^{1/4}
-(2|Q_0|+\delta_{Q_0,0})\frac1{\sqrt{2\pi t/\tau}}\\
+(2Q_0^2+1)\frac{\Gamma(\frac34)}{4\pi}\left(\frac{2}{t/\tau}\right)^{3/4}+O(t^{-5/4})\,.
\end{multline}
As in the perfectly transmitting case, the leading
asymptotic behavior of $P(Q_0;t)$ is the same for any fixed $Q_0$, but now $P(Q_0;t)$ falls off as $\sim t^{-1/4}$ instead of $\sim t^{-1/2}$, 
and similarly, the width of the distribution also scales as $(t/\tau)^{1/4}$ (see also next subsection) rather than  $(t/\tau)^{1/2},$ as found in the transmissive case. However, for any $t$ the distribution is nontrivial, as reflected by the quite unusual non-analytic $\propto |Q_0|$ term in Eq.~\eqref{eq:PQrepas}.
We compared this result with the numerical evaluation of the discrete sum \erf{eq:PQr} and found good agreement.

We can derive another large-time property by replacing ${\cal P}(s)$ by its long time asymptotic form ${\cal P}(s)\approx e^{-s^2/(2t/\tau)}/\sqrt{2\pi t/\tau}$ 
and approximating the binomial distribution by a Gaussian (valid for large $|s|$), yielding
\be
\label{eq:PQras}
P^\text{(r)}(Q_0;t)\approx \int_{0}^\infty \ud s \frac{e^{-s^2/(2t/\tau)}}{\sqrt{2\pi t/\tau}} \sqrt{\frac2{\pi s}} e^{-Q_0^2/(2s)}\,.
\ee
This approximate expression is compared to the exact result \erf{eq:PQr} in Fig. \ref{fig:PQ}(a) for $t/\tau=75.$ Apart from $Q_0=0,$ an excellent agreement is found.

In fact, Eq. \erf{eq:PQras} gives the \emph{time-independent} distribution of the scaled transported charge $\tilde Q_0 \equiv (t/\tau)^{-1/4} Q_0$ at large times. Indeed, after a trivial change of integration variables we obtain
\begin{multline}
\label{eq:PQrtyp}
P^\text{(r)}(\tilde Q_0) =(t/\tau)^{1/4} P\left(Q_0=(t/\tau)^{1/4} \tilde Q_0 ; t\right)\\
=\frac1\pi \int_{-\infty}^\infty \ud u \,e^{-u^4/2-\tilde Q_0^2/(2u^2)}\,.
\end{multline}
As we will show in the next subsection, the variance of $P(Q_0;t)$ scales as $\sim t^{1/2}$ for large times, so Eq. \erf{eq:PQrtyp} describes the distribution of the \emph{typical} fluctuations of $Q_0$ which are \emph{non-Gaussian}.
Expression \erf{eq:PQras} already appeared in Ref. \cite{Altshuler2006} and was also found independently in Ref. \cite{Krajnik2022} for the distribution of the integrated charge current in a cellular automaton model involving two species of hard-core particles having opposite charges. This model can be viewed as the discretized version of a perfectly reflective semiclassical system in which particles move at a fixed, constant speed. The authors of Refs. \cite{Krajnik2022,Krajnik2022c} argued that Eq. \erf{eq:PQrtyp} gives the universal equilibrium distribution of typical current fluctuations in systems obeying the single file property.

We studied the generic case with finite transmission rate numerically, both via the semiclassical simulation and within the SEP approximation. In the latter the distribution function was computed by a numerical Fourier transformation of the generating function calculated by Eq. \erf{eq:corr2} and using the exact SEP result for the charge average. The results are shown in Fig. \ref{fig:PQ}(b) at different times for temperature $T=0.02Mc^2$ and $\gamma=0.75$ where the mean transmission probability is $\pt\approx0.1.$ There is an excellent agreement between the semiclassical simulations and the SEP approximation. In Fig. \ref{fig:PQ}(a) the distributions in the two extreme limits are plotted together with the results obtained in this generic case for the same time. Naturally, the distribution is broader for greater transmission probability but remains very different from the purely transmissive distribution whose width scales differently with time.

\subsection{Moments and cumulants}

We now discuss the moments $\langle Q_0^n\rangle$ and cumulants $\langle Q_0^n\rangle_\text{c}$ 
of the distribution $P(Q_0;t)$.
The moments of the transferred charge (integrated topological current) distribution 
can be obtained by differentiating $C_\eta(0,t)$ with respect to $\bar\eta$ and then setting $\bar\eta$ to 
zero. The cumulants can be derived in the same way from $\ln \,C_\eta(0,t).$ 

In the totally transmitting case, we find from Eq. \erf{eq:autoCtr} that 
the even moments scale as
\be
\langle Q_0^n\rangle^\text{(tr)}\sim \left( {t}/{\tau}\right)^{n/2}\,,
\ee
while all odd moments vanish. All even cumulants are equal, 
\be
\langle Q_0^n\rangle^\text{(tr)}_\text{c} = \frac{t}\tau\,,
\ee
showing that the distribution 
never becomes Gaussian, although for large times all the moments approach the Gaussian moments in the leading 
order, so the Gaussian becomes an excellent approximation of the true distribution.

In the perfectly reflecting case, 
the integral representation \erf{eq:corr} cannot be used because differentiating the integrand leads to a divergent integral. 
We use instead Eq. \erf{eq:PQr} to write
\be
\label{eq:Qn}
\bigl\langle Q^n_0(t)\bigr\rangle^\text{(r)} = \sum_{s=-\infty}^\infty {\cal P}(s) \bigl\langle\Delta Q(s;t)^n\bigr\rangle\,,
\ee
where $s=\kr-\kl$, and the moments of $P_s(\Delta Q;t)$ appear.
In this case we can use the well-known moments of the binomial distribution to evaluate $ \langle\Delta Q(s;t)^n\rangle $, 
and the final moments can be given in a closed form in terms of modified Bessel functions. For example, the second moment reads as
\begin{multline}
\langle Q_0^2(t)\rangle^\text{(r)}= \sum_{s=-\infty}^\infty e^{-t/\tau} I_{|s|}(t/\tau) |s| \\
= t/\tau e^{-t/\tau}\bigl[I_0(t/\tau)+I_1(t/\tau)\bigr]
\,.
\end{multline}
Some higher moments are listed in Appendix \ref{app:FCSrep}. 
In light of the general case treated below, it is instructive to investigate the large time behavior of the moments in this reflective 
limit by approximating $\mathcal{P}(s)$ by its asymptotic Gaussian form\footnote{The resulting expression also follows from evaluating the integral in Eq. \erf{eq:corr} in the saddle point approximation.} and approximating the sum over $s$ by an integral,
which leads to
\bes
\label{eq:Qnrefl}
\begin{align}
\langle Q_0^2(t)\rangle^\text{(r)}&\approx \!\int\!
\ud s \frac{e^{-s^2/(2t/\tau)}}{\sqrt{2\pi t/\tau}}|s| = \sqrt{\frac2\pi\frac{t}\tau}\,,\\
\langle Q_0^4(t)\rangle^\text{(r)}&\approx \!\int\!
\ud s \frac{e^{-s^2/(2t/\tau)}}{\sqrt{2\pi t/\tau}}(3s^2-2|s|)
=3\frac{t}\tau-\sqrt{\frac8\pi\frac{t}\tau}\,,
\end{align}
\esu
and so on. 
The even moments thus grow for large times as 
\be
\bigl\langle Q_0(t)^n\bigr\rangle^\text{(r)}\sim (t/\tau)^{n/4}
\ee
in the reflective limit. This must be contrasted to the  
$\langle Q_0(t)^n\bigr\rangle^\text{(t)}\sim (t/\tau)^{n/2}$ behavior in the transmissive case. 
The same scaling holds for the cumulants; 
in stark contrast to the transmissive limit, now the leading contributions do not cancel.

The kurtosis (``tailedness'') of the distribution is given by
\be
K = \frac{\langle Q_0(t)^4\rangle}{\langle Q_0(t)^2\rangle^2}\approx \frac{3\pi}2-\sqrt{\frac{2\pi}{t/\tau}}\,.
\ee 
It approaches $3\pi/2\,\approx 4.71$ which is clearly larger than 3, the kurtosis of the normal distribution.

Away from the  special cases of perfect transmission and perfect reflection, 
for small transmission rates, we can again make use of the mapping to the SEP and the exact moments of 
$P(\Delta Q(s;t))$ obtained in Ref.~\cite{Imamura2021}. Although Eq.~\eqref{eq:Qn} is not valid in general, 
it holds in the  SEP approximation, where the distribution of the transferred charge depends only 
on $s=\kr-\kl.$
The first two nontrivial moments are 
\bes
\begin{align}
\langle \Delta Q(s;\hat t)^2\rangle^\text{SEP}  &= |s| + 2 J_1(s,\hat t)\,,\\
\langle \Delta Q(s;\hat t)^4\rangle^\text{SEP}  &= (3s^2-2|s|)+4(3|s|+2) J_1(s,\hat t)\nonumber\\
&\phantom{=}+ 12 J_2(s,\hat t)\,,
\end{align}
\esu
where $J_k$ are given by $k$-fold integrals (see Appendix \ref{app:SEP}). We can see that the moments of the binomial distribution valid for perfect reflections get modified in a time dependent manner. 

The asymptotic behavior of $J_k(s,\hat t)$ is also known in the limit $|s|,t\to\infty,$ $\xi=-s/\sqrt{4t}$ fixed \cite{Imamura2021} (c.f. Appendix \ref{app:SEP}). Plugging these in Eq. \erf{eq:Qn} and approximating the sum over $s$ by an integral over $\xi$ yields
\bes
\label{Qnasym}
\begin{align}
\langle Q_0(t)^2\rangle^\text{SEP} &\approx \sqrt{\frac{2\alpha}\pi}\int_{-\infty}^\infty \ud\xi\,   e^{-2\xi^2\alpha} \sqrt{\hat t}A(\xi)\nonumber\\
&= \sqrt{\frac2\pi}\sqrt{\frac{t}{\tau}}\sqrt{1+2\alpha}\,,\\
\langle Q_0(t)^4\rangle^\text{SEP} &\approx 
\sqrt{\frac{2\alpha^3}\pi}\frac{t}{\tau}\int_{-\infty}^\infty \ud\xi\,   e^{-2\alpha\xi^2} 12 A(\xi)^2 \nonumber\\
&\phantom{=}-\sqrt{\frac8\pi}\sqrt{\frac{t}{\tau}}(3\sqrt{1+\alpha}-2\sqrt{1+2\alpha})\,,
\end{align}
\esu
where $A(\xi)$ is defined below Eq. \erf{eq:SEPas} and $\alpha$ is defined in Eq. \erf{alphadef} as the proportionality factor between 
the SEP time $\hat t$ and dimensionless physical time $t/\tau.$

We can see that the powers of $t$ are the same as for the totally reflective case, only their 
coefficients get modified. The statistical properties of charge transfer 
in the general case are thus closer to the completely reflective limit. Indeed, the
 last integral approaches $3\sqrt{\pi/(2\alpha^3)}$ for $\alpha\propto p_T \to0$ 
from above, so in the reflective limit, $\pt\to0$, we recover the reflective 
results in Eqs. \erf{eq:Qnrefl}.  The kurtosis is larger than in the completely reflective case.

\begin{figure}[t]
\centering
\includegraphics[width=0.49\textwidth]{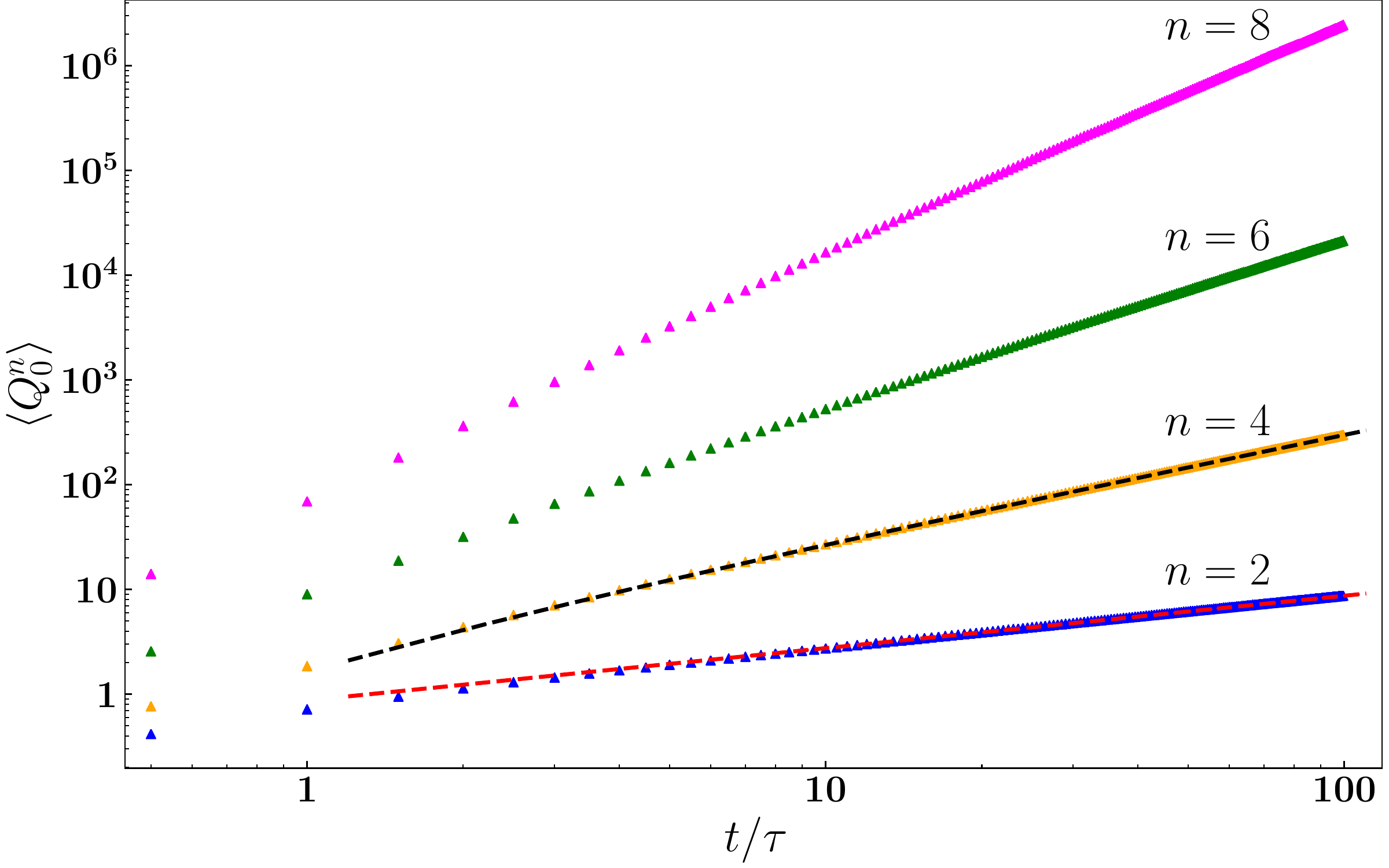}
\caption{Moments of the transferred charge $\langle Q_0^n\rangle$ for $n=2,4,6,8$ as functions of time at $\gamma=0.8$ and $T=0.1Mc^2.$
Symbols denote results of 
the semiclassical simulation, asymptotic results \erf{Qnasym} are plotted as dashed lines. \label{fig:moments} }
\end{figure}

 In Fig. \ref{fig:moments}, we show the Monte Carlo results for the first four even moments 
 on a log-log scale, together with the SEP asymptotics \erf{Qnasym} for the first two. 
 We checked that, although the correction with respect to the reflective results are small in the 
 small $p_T$ regime, Eqs.~\erf{Qnasym} give better agreement than Eqs.~\erf{eq:Qnrefl},
  valid in the totally reflective case.

We note that the fact that the  cumulants scale with different powers of $t$ means that they cannot 
be obtained from a scaled cumulant generating function. According to Eq. \erf{eq:SEPasym}, the 
latter scales as $\sim\sqrt{t}$ but this does not carry over to the cumulants. A similar 
anomalous behavior was recently observed in Refs. \cite{Krajnik2022,Krajnik2022c,Krajnik2022b,Gopalakrishnan2022}.

\section{Topological charge density correlators}
\label{sec:qq}

The topological charge density operator is given by
\be
\varrho_q(x) = \frac\gamma{2\pi} \partial_x\Phi(x)\,.
\ee
Its correlation functions,
\be
R(x,t)\equiv \langle \varrho_q(x,t)\varrho_q(0,0)\rangle\,,
\ee
can be obtained \cite{Damle2005} from the vertex operator correlation function by extracting the second order term in its series expansion in $\eta$ and then differentiating twice with respect to $x.$

In the purely transmissive case, from Eq. \erf{Ctr} we obtain
\be
\left.\frac12 \frac{\p^2 C^\text{(tr)}(x,t)}{\p\eta^2}\right|_{\eta=0} = -\frac{\pl+\pr}2 \left(\frac{2\pi}\gamma\right)^2\,.
\ee
Using (see Appendix \ref{app:qq})
\be
\p_x^2 (\pl+\pr) = \p_x^2 \,\rho\!\int_{-c}^{c}\! \ud v f(v) |x-vt| 
=
\frac{2\rho}t f(x/t)\,,
\ee
we obtain
\be
R^\text{(tr)}(x,t) = \rho\frac{f(x/t)}t\,,
\ee
where $f(v)$ is given in Eq. \erf{fv}. At low temperatures, we recover the free density-density correlation function,
\be
R^\text{(tr)}(x,t) \approx \frac{M}{2\pi t}e^{-Mc^2/T}e^{-M x^2/(2t^2)}\,.
\ee
In particular, the autocorrelation function is
\be
R^\text{(tr)}(0,t) = \frac{M}{2\pi t}e^{-Mc^2/T} \approx \frac{\rho^2}{\pi t/\tau}\,,
\ee
where the $1/t$ behavior emerges in the limit where the extension of the topological 
charges tends to zero.

In the purely reflective case, the calculation is more complicated. We differentiate twice the integrand of Eq. \erf{eq:Cr} with respect to $x$, and then with respect to $\bar\eta.$ The resulting integral expression is shown in Appendix \ref{app:qq}. It can be evaluated for the autocorrelation function with the result
\begin{multline}
\label{qqR}
R^\text{(r)}(0,t) \\
=\frac{M}{2\pi t}e^{-Mc^2/T} e^{-t/\tau} I_0(t/\tau)
+\rho^2 e^{-t/\tau} \frac{I_0(t/\tau)+I_1(t/\tau)}2\\
\approx \frac12\rho^2 e^{-t/\tau} \left(\frac2{\pi t/\tau} I_0(t/\tau)+I_0(t/\tau)+I_1(t/\tau)\right)\,.
\end{multline}
For short times $t\ll\tau$ we recover the transmissive result,
\be
R^\text{(r)}(0,t) \approx  \frac{M}{2\pi t}e^{-Mc^2/T} e^{-t/\tau} + \frac12\rho^2+\dots\,,
\ee
while the large time asymptotics is of the diffusive form \cite{Damle2005}
\be
R^\text{(r)}(0,t) \approx \frac{\rho^2}{\sqrt{2\pi t/\tau}} + O[(t/\tau)^{-3/2}]\,.
\ee

\begin{figure}[t]
\centering
\includegraphics[width=0.99\columnwidth]{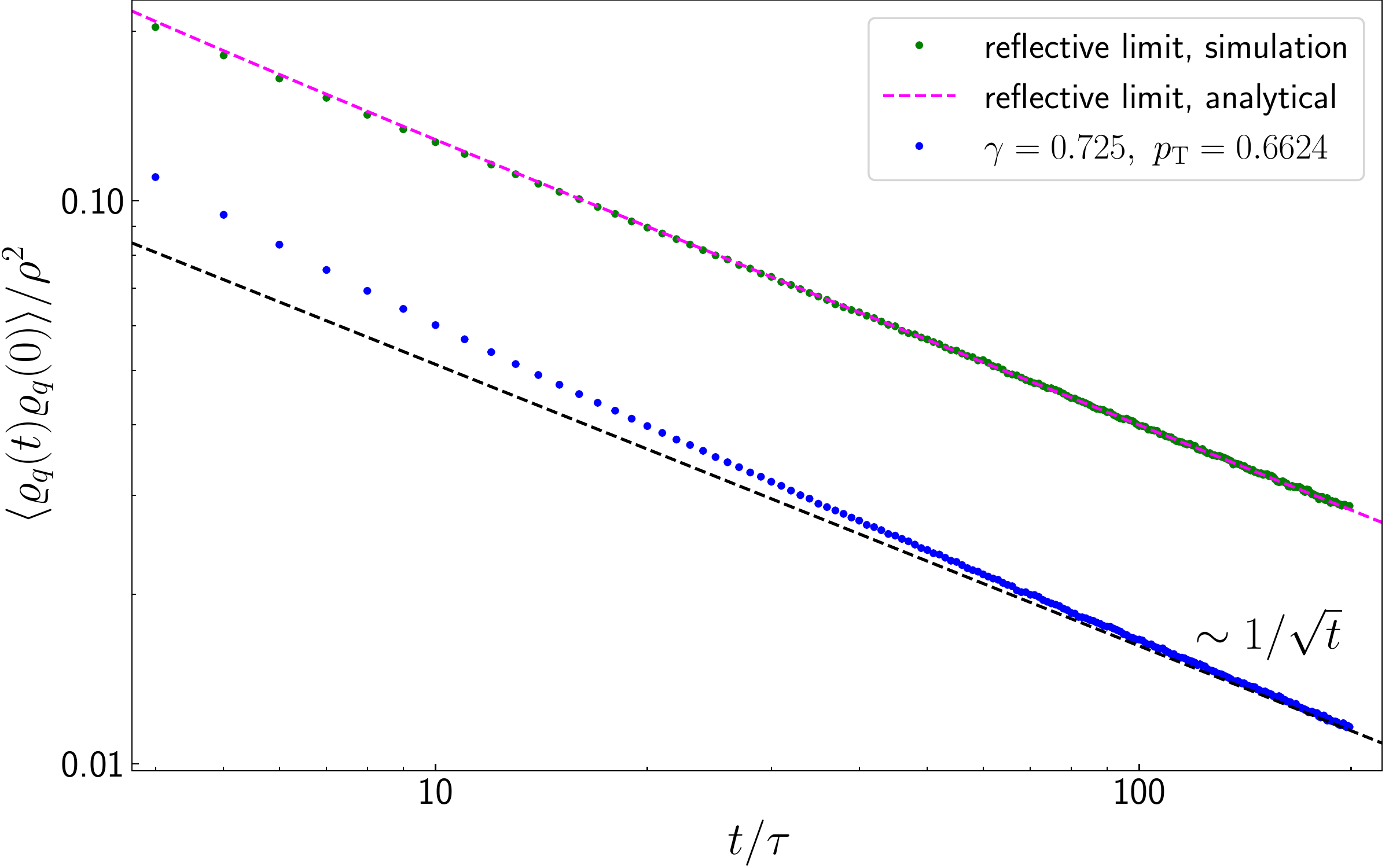}
\caption{Topological charge density autocorrelation function normalized by the particle density squared on a log-log scale. 
The upper curves correspond to the purely reflective case, while the lower curves refer to $\gamma=0.725,$ the temperature is set to $T=0.1Mc^2.$
The purely reflective simulation results are plotted in green dots, while the magenta dashed line shows the analytic result \erf{qqR}. The blue dots represent the simulation results for $\gamma=0.725$, and the black dashed line indicates the large time $\sim t^{-1/2}$ asymptotics. \label{fig:density}}
\end{figure}

We show our semiclassical numerical results for the charge density autocorrelation function in Fig. \ref{fig:density}. In the numerical calculations, 
shown as dots, the charge density was measured in finite intervals. To obtain better precision and to reduce noise, we 
varied the length of the intervals and used shorter intervals at short times.

The upper (green) data points were obtained for the totally reflective case and are seen to match the analytic expression \erf{qqR} shown in magenta dashed line. The lower (blue) dots show our numerical results for $\gamma=0.725$ where both transmissive and reflective collisions are present. The asymptotic behavior is found to be consistent with the diffusive $\sim t^{-1/2}$ form also in this case, providing evidence for the 
generic diffusive nature of charge transport.

\section{Conclusions}
\label{sec:concl}

In this work, we studied  topological charge dynamics of generic, non-integrable systems in the sine--Gordon family 
at low temperatures. The systems in this class possess a natural phase-like field and 
associated long-lived, gapped topological ``kink'' excitations carrying topological charges. 
These models arise naturally as effective low-energy descriptions of various 1D quantum 
systems via bosonization.

We investigated the equilibrium, finite temperature 
correlation function of the vertex operators,  $\big\langle e^{i\eta \Phi(x,t)}e^{-i\eta \Phi(0,0)}\big\rangle_T$, 
the distribution function 
of transferred topological charge $P(\Delta Q;t)$ and its cumulants as well as
the correlation function of the topological charge density $\varrho_q(x,t)$ using
the hybrid semiclassical method developed in Ref.~\cite{Moca2016}.
 This approach is
formulated in terms of the physical massive kink particles, and is supposed to be valid at 
 temperatures small compared to the energy gap (the kink mass). It allows us to go
  beyond the original semiclassical approach that assumes perfectly reflective scatterings~\cite{Sachdev1996,Sachdev1997,Damle2005}, 
 and to study the effect of transmissive collisions, present at any finite temperature. We found that, 
 in contrast to the prediction of the original semiclassical approach, the vertex operator correlation 
 function does not decay diffusively in time but follows a \emph{stretched exponential} decay, 
 multiplied by a power law (c.f. Eq. \erf{eq:SEPasym}). This generic result is different  from 
 the analytically tractable purely reflective (diffusive power law) and purely transmissive 
 (ballistic exponential) cases. 

We substantiated these results  analytically in the weakly transmissive limit, by approximating the system 
and mapping it to the simple symmetric exclusion process. The decay of correlations comes from two 
sources: the random motion of the domains bordered by the kinks (see Fig. \ref{fig:domains}) giving rise to a slow, 
diffusive decay, and from the domains changing their charges due to  transmissive collisions, 
yielding an exponential decay. However, since the charge of a given domain  propagates diffusively, 
due to the random domain wall collisions,
the exponent scales as $\sim\sqrt t$ instead of being linear in time.

Even though the SEP approach is well justified  only in the case of rare transmissive collisions, our  numerical semiclassical  simulations showed that the obtained time dependence holds much more 
generally, even for relatively high average transmission probabilities.

While the generic (finite transmission) vertex correlation functions are very different from the
ones computed in the  reflective limit, the autocorrelation function of the topological charge density is found to 
 behave generically diffusively, just like in the purely reflective case. This is somewhat expected: as the 
probability of reflective collisions is finite, at some point the charges get reflected, implying a diffusive, 
Brownian motion. It is interesting to note, however, that even though the charge dynamics is diffusive, 
the vertex operator correlation function behaves very differently in the purely reflective and in the generic 
case, so it appears to be  a more  sensitive probe of the dynamics.

Our approach also allowed us to study the statistics of transferred topological charge, 
$P(\Delta Q;t)$, as a function of time. We derived analytical results in the purely reflective 
and transmissive limits, and have shown that this function as well as the moments 
and cumulants of the transferred charge behave strikingly differently 
in these limits.
In the transmissive (ballistic) limit, the moments $\langle \Delta Q(t)^n\rangle^\text{(tr)}$ scale as $\sim t^{n/2}$
(for $n$ even), while in the purely reflective limit, $\langle \Delta Q(t)^n\rangle^\text{(r)}\sim t^{n/4}$.
The origin of the difference is very simple; in the transmissive limit, $N\propto t$ topological 
charges of random sign  are transferred through a given point, yielding  $\Delta Q \sim \sqrt{N}\sim t^{1/2}$, and 
  $\langle \Delta Q(t)^n\rangle^\text{(tr)}\sim  t^{n/2}$. 
In the reflective limit, however, topological charges move diffusively, and the number of charges 
passing a given point scales as $N \sim \sqrt{t}$ only. The sum of these  $\sim \sqrt{t}$
random charges then scales as  $\Delta Q \sim t^{1/4}$, yielding the observed scaling, 
$\langle \Delta Q(t)^n\rangle^{(r)} \sim  t^{n/4}$. 

In the transmissive case, the cumulants scale ballistically, $\sim t,$ whereas in the purely reflective case they follow the moments and scale non-uniformly. This implies that the scaled cumulant generating function is anomalous as it does not generate faithfully the scaled cumulants \cite{Krajnik2022c}. In the reflective limit, we recover 
the same ``universal'' distribution of typical fluctuations as a recent, independent work \cite{Krajnik2022} 
that studied a quantum cellular automaton involving two kinds of charges, which can be viewed as a 
discretized version of our model. 

Since in the generic case a small backscattering is always relevant,
 the asymptotic behavior of moments and cumulants 
is  similar to what we find in the reflective case, though small corrections appear and 
the prefactors obviously differ.

Even though we used the sine--Gordon S-matrix for convenience, our main results do 
not depend on this choice. Importantly, whether the S-matrix satisfies the Yang--Baxter equation 
played no role in our study that focused on generic, non-integrable versions of the sine--Gordon theory. 
We also neglected some coherent parts of the scattering histories. These coherent  contributions become
 important in the integrable case, when the S-matrix obeys the Yang--Baxter equation, and the exact 
 eigenstates of the system can be written in terms of ballistically propagating quasiparticles, 
 the so-called magnons. This representation is used in the
 generalized hydrodynamics of the integrable sine--Gordon model \cite{Bertini2019} predicting ballistic charge transport, 
 exponentially decaying vertex operator autocorrelation functions, and uniformly scaling cumulants for the 
 integrated topological current \cite{inprep}.
  
\emph{Physical} systems belonging to the sG family are, however, 
not integrable, and the S-matrix typically does \emph{not} satisfy the Yang--Baxter relations. 
In this case integrability-related  interference features are expected to be unimportant or 
 to appear in  a restricted region.  We therefore expect our results to describe the dynamics  in 
these generic systems.

~\vskip1cm
\begin{acknowledgments}
We thank Benjamin Doyon and G\'abor Tak\'acs for illuminating discussions and  Toma\v{z} Prosen for drawing our attention to the recent works \cite{Krajnik2022,Krajnik2022b,Krajnik2022c}.
This research was supported by the National Research, Development and Innovation Office - NKFIH through research 
grants Nos. K134983, K138606, and SNN139581, and within the Quantum National Laboratory of Hungary 
program (Project No. 2017-1.2.1-NKP- 2017-00001). M.K. acknowledges support by a ``Bolyai J\'anos'' grant of the HAS and by the \'UNKP-21-5 new National Excellence Program of the Ministry for Innovation and Technology from the source of the National Research, Development and Innovation Fund.
\end{acknowledgments}



\appendix

\onecolumngrid

\section{Exact results for the SEP}
\label{app:SEP}

In this appendix we quote some exact results relevant to our work derived in Ref \cite{Imamura2021} for the simple symmetric exclusion process. Following this reference, let us denote by $N(0,\hat t)$ the integrated particle current at the origin, i.e., the total number of particles that have jumped from site $1$ to $0$ minus the total number of particles that have jumped from $0$ to $1$ in SEP time $\hat t$ (note the somewhat unusual convention for the sign of the current). Let us also define the quantity $N(s,\hat t)$ for $s\neq0$ such that 
\be
N(s,\hat t) = 
\begin{cases}
N(0,\hat t)& +  \sum_{r=1}^s n_r(\hat t)\qquad \qquad s>0\,,\\
N(0,\hat t)& - \sum_{r=-|s|+1}^0 n_r(\hat t)\qquad s<0\,,
\end{cases}
\ee
where $n_r(\hat t)\in\{0,1\}$ is the site occupation at site $r$ at time $\hat t.$ 
This is closely related to the topological charges in domains at different times that are $s$ kinks apart from each other. The precise relation is 
\be
\label{eq:QNrel}
Q(s,\hat t)-Q(0,0) = 2 N(s,\hat t) -s\,.
\ee
To see this, note that the topological current is (minus) twice of the SEP particle current, because a particle jump corresponds to swapping two opposite charges. When we move along a vertical (time-like) segment in the calculation of $\Delta Q(s,\hat t),$ we need the negative of the current, so the sign is correct. Along the horizontal (space-like) segment, we need to add the charges, which using $q_j=2n_j-1$ gives the above relation.

In the work \cite{Imamura2021}, an exact result was derived for the characteristic function of the probability distribution of $N(s,\hat t)$ for arbitrary $s$ and $\hat t$ for a Bernoulli initial condition with a step-like density profile. Here we specify to the case where the left and right densities are both $1/2.$
The characteristic function is written in terms of a Fredholm determinant:  
\be
\left\langle e^{\lambda N(s,\hat t)}\right\rangle = \left\langle e^{\lambda/2\, \Delta Q(s,\hat t)}\right\rangle e^{\lambda s/2}=
\mathrm{det}\Big(1+\sinh^2(\lambda/2)\,K_{s,\hat t}\Big)_{L^2(C_0)}\cdot \left(\frac{1+e^{\mathrm{sgn}(s)\lambda}}2\right)^{|s|}\,,
\ee
where the kernel is
\be
K_{s,\hat t}(\xi_1,\xi_2) = \frac{\xi_1^{|s|}e^{(\xi_1+1/\xi_1-2)\hat t}}{\xi_1\xi_2+1-2\xi_2}\,,
\ee
and $C_0$ is a contour around the origin small enough so the poles of the kernel lie outside of it. 

For $s\ge0,$ the moments of the distribution can be expressed as
\be
\vev{N(s,\hat t)^n} = \sum_{k=0}^n m_{n,k}(s) J_k(s,\hat t)\,,
\ee
where
\be
J_k(s,\hat t) = \int_{C_0}\dots  \int_{C_0}\prod_{i<j} \frac{\xi_i-\xi_j}{\xi_i\xi_j+1-2\xi_j}\prod_{i=1}^k\frac{\xi_i^s e^{(\xi_i+1/\xi_i-2)\hat t}}{(1-\xi_i)^2}\,,
\ee
and the coefficients $m_{n,k}$ are given implicitly by
\be
\sum_{n=0}^\infty \frac{\lambda^n}{n!} m_{n,k}(s) = \frac{\sinh^{2k}(\lambda/2)}{k!}\left(\frac{e^\lambda+1}2\right)^s\,.
\ee
 For example,
 \begin{align}
 m_{0,0}&=1\,,\quad m_{1,0}=\frac{s}2\,,\quad m_{2,0}=\frac{s(1+s)}4\,,\quad m_{3,0}=\frac{s^2(3+s)}8\,,\quad m_{4,0}=\frac{s(1+s)(s^2+5s-2)}{16}\,,\\
 m_{0,1}&=0\,,\quad m_{1,1}=0\,,\quad m_{2,1}=\frac12\,,\quad m_{3,1}=\frac{3s}4\,,\quad m_{4,1}=\frac{3s^2+3s+2}4\,.
 \end{align}

The long-time asymptotic behavior of the characteristic function in the limit $\hat t\to\infty,$ $s\to\infty,$ $\xi\equiv-s/\sqrt{4\hat t}$ fixed was also derived in Ref. \cite{Imamura2021}:
\be
\label{eq:SEPres}
\left\langle e^{\lambda N(s,\hat t)}\right\rangle \sim e^{-\sqrt{\hat t} \mu(\xi,\lambda)}\,,
\ee
where
\be
\mu(\xi,\lambda) = \sum_{n=1}^\infty \frac{\left(-\sinh^2(\lambda/2)\right)^n}{n^{3/2}} A(\sqrt{n}\xi)+\lambda \xi
\ee
with
\be
A(y) = \frac{e^{-y^2}}{\sqrt\pi} + y\, \mathrm{erf}(y)\,.
\ee
In the same asymptotic limit, the long-time behavior of the moments can be obtained based on the asymptotic behavior of the $J_n(s,\hat t)$ functions:
\begin{align}
J_1(s,\hat t)&\sim \sqrt{\hat t} (A(\xi)-|\xi|)\,,\\
J_2(s,\hat t)&\sim \hat t (A(\xi)-|\xi|)^2-\sqrt{\hat t/2} \left(A(\sqrt2\xi)-\sqrt2|\xi|\right)\,.
\end{align}

We now translate these results to our language. In our case, $\lambda=2i\bar\eta,$ and
\be
\left\langle e^{i\bar\eta\Delta Q(s,\hat t)}\right\rangle =
\mathrm{det}\Big(1-\sin^2(\bar\eta)\,K_{s,\hat t}\Big)_{L^2(C_0)}\cdot (\cos\bar\eta)^{|s|}\,.
\ee
Notice the $(\cos\bar\eta)^{|s|}$ factor corresponding to the binomial distribution at $\hat t=0.$ This is the purely reflective result which corresponds no SEP dynamics, i.e., $\hat t=0.$  

The moments of $\Delta Q(s,\hat t)=2N(s,\hat t)-s$ are obtained as linear combinations of moments of $N(s,\hat t).$ The odd moments obviously vanish, and the even moments are even in $s$:
\begin{align}
\vev{\Delta Q(s,\hat t)^2} &= |s| + 2 J_1(s,\hat t)\,,\\
\vev{\Delta Q(s,\hat t)^4} &= (3s^2-2|s|)+4(3|s|+2) J_1(s,\hat t) + 12 J_2(s,\hat t)\,.
\end{align}
Finally, the asymptotic expression \erf{eq:SEPas} in the main text follows simply from Eqs. \erf{eq:QNrel} and \erf{eq:SEPres}.

\section{Full counting statistics in the perfectly reflecting case}
\label{app:FCSrep}

We start from
\be
P(Q_0;t) = \int_0^{2\pi}\frac{\ud\bar\eta}{2\pi}e^{-i\bar\eta Q_0} C_\eta(0,t)\,.
\ee
Using Eq. \erf{eq:Cr}
\be
\label{eq:PQint}
P(Q_0;t) =
\int_0^{2\pi} \frac{\ud\phi}{2\pi} 
F(Q_0,\phi)
e^{-\sin^2(\phi/2)t/\tau}\,,
\ee
where
\be
F(Q_0,\phi) =  \int_0^{2\pi}\frac{\ud\bar\eta}{2\pi}e^{-i\bar\eta Q_0}\frac{1-\cos^2\bar\eta}{1-2\cos\bar\eta\cos(\phi)+\cos^2\bar\eta}\,.
\ee
Alternatively, $F(Q_0,\phi)$ can be obtained from Eq. \erf{eq:corr} by taking the Fourier transform of $\vev{e^{i\bar\eta\Delta Q_s(t)}}=(\cos\bar\eta)^{|s|}$ to obtain
\be
P_s^{\text{(r)}}(Q_0) =  \int_0^{2\pi}\frac{\ud\bar\eta}{2\pi}e^{-i\bar\eta Q_0} (\cos\bar\eta)^{|s|} = \frac1{2^{|s|}}\binom{|s|}{\frac{Q_0+|s|}2}\,,
\ee
and then calculating the sum
\be
F(Q_0,\phi) = \sum_{s=-\infty}^\infty \frac1{2^{|s|}}\binom{|s|}{\frac{Q_0+|s|}2}\ e^{-is\phi}\,.
\ee
In practice, the sum runs over $|s|\ge |Q_0|$ in steps of two.
The first few examples are
\begin{align}
F(0,\phi) &= -1+\sqrt{1+\frac1{|\sin\phi|}} \,,\\
F(\pm1,\phi)&=
\frac{\cos(3\phi/2)+\sin(3\phi/2)}{\sqrt{\sin\phi}}-2\cos\phi
\,.
\end{align}
The general result for any $Q_0\neq0$ is
\be
F(Q_0,\phi) = \frac{e^{-i\phi |Q_0|}}{2^{|Q_0|}}\,_2F_1\Big(\frac{1+|Q_0|}2,1+\frac{|Q_0|}2;1+|Q_0|,e^{-2i\phi}\Big)
+\frac{e^{i\phi |Q_0|}}{2^{|Q_0|}}\,_2F_1\Big(\frac{1+|Q_0|}2,1+\frac{|Q_0|}2;1+|Q_0|,e^{2i\phi}\Big)\,,
\ee
where $_2F_1(a,b;c;z)$ is the hypergeometric function. Its Taylor expansion around $\phi=0$ reads
\be
F(Q_0,\phi) = \frac1{\sqrt\phi}-2Q_0+\frac{2Q_0^2+1}2\sqrt\phi+O(\phi^{3/2})\,,
\ee
Plugging this into Eq. \erf{eq:PQint} and evaluating the integral in the saddle point approximation we obtain Eq. \erf{eq:PQrepas}.

The first few moments of the transferred charge can be given in a closed form using Eq. \erf{eq:PQr} and the moments of the binomial distribution:
\begin{align}
\big\langle Q_0(t)^2\big\rangle&=\sum_{s=-\infty}^\infty e^{-t/\tau} I_{|s|}(t/\tau) |s| = 
 t/\tau e^{-t/\tau}\bigl[I_0(t/\tau)+I_1(t/\tau)\bigr]\,,\\
\big\langle Q_0(t)^4\big\rangle&=\sum_{s=-\infty}^\infty e^{-t/\tau} I_{|s|}(t/\tau) (3s^2-2|s|) = 3t/\tau-2 t/\tau e^{-t/\tau}\bigl[I_0(t/\tau)+I_1(t/\tau)\bigr]\,,\\
\big\langle Q_0(t)^6\big\rangle&=\sum_{s=-\infty}^\infty e^{-t/\tau} I_{|s|}(t/\tau) (15|s|^3-30s^2+16)  \\
&=-30t/\tau+(30t/\tau+31) t/\tau e^{-t/\tau}I_0(t/\tau)+I_1(t/\tau)+(30t/\tau+16) t/\tau e^{-t/\tau}I_1(t/\tau)\,.
\end{align}

\section{Topological charge density correlation function}
\label{app:qq}

In this Appendix, we detail the calculation of the charge correlators. These can be obtained from the correlation function \erf{eq:C} by differentiating it with respect to $x$ and $\eta.$ Let us consider the correlator of operators inserted at $x_1$ and $x_2.$ Differentiating it with respect to these coordinates we find
\be
\p_{x_1}\p_{x_2} \vev{e^{i\eta \Phi(x_1,t)}e^{-i\eta \Phi(x_2,0)}} = 
\eta^2 \vev{\p_{x1}\Phi(x_1,t)e^{i\eta \Phi(x,t)}e^{-i\eta \Phi(0,0)}\p_{x2}\Phi(x_2,0)}\,.
\ee
Taking two $\eta$-derivatives and setting $\eta=0$ we arrive at
\be
\frac12\p_\eta^2\p_{x_1}\p_{x_2} \left.\vev{e^{i\eta \Phi(x_1,t)}e^{-i\eta \Phi(x_2,0)}}\right|_{\eta=0} =
 \vev{\p_{x1}\Phi(x_1,t)\p_{x2}\Phi(x_2,0)} = \left(\frac{2\pi}\gamma\right)^2\vev{\varrho_q(x_1,t)\varrho_q(x_2,0)}\,.
\ee
Due to translational invariance, the correlation function depends on $x_1-x_2$ only, so $\p_{x_2}=-\p_{x_1}.$ So we can reset $x_1=x$, $x_2=0,$ and then
\be
\vev{\varrho_q(x,t)\varrho_q(0,0)} = - \left.\frac12\frac{\p^2}{\p\bar\eta^2}\frac{\p^2}{\p x^2} C_\eta(x,t)\right|_{\bar\eta=0}\,,
\ee 
where we used $\bar\eta=2\pi\eta/\gamma.$

The correlation function $C_\eta(x,t)$ depends on x through the functions $\pr(x,t)$ and $\pl(x,t).$ Their $x$-derivatives are
\begin{align}
\p_x\pr = \phantom{-}&\rho \int_{-c}^c\ud v f(v)\Theta(x/t-v) \,,\\
\p_x\pl = -&\rho \int_{-c}^c\ud v f(v)\Theta(v-x/t) \,,
\end{align}
and
\be
\p^2_x\pr = \p^2_x\pl = \frac\rho{t} f(x/t)\,.
\ee
Taking the derivatives of Eq. \erf{eq:Cr} we obtain
\begin{multline}
\label{eq:qq}
\vev{\varrho_q(x,t)\varrho_q(0,0)} = \frac12\int_0^{2\pi}\frac{\ud\phi}{2\pi}\frac1{2\sin^2(\phi/2)}e^{-2\sin^2(\phi/2)(\pr+\pl)+i\sin\phi(\pr-\pl)}\\
\left[4\sin^2(\phi/2)\frac\rho{t}f(x/t)-\left(2\sin^2(\phi/2)\rho \int \ud v f(v)\mathrm{sgn}(x-vt)-i\rho\sin\phi\right)^2  \right]\,.
\end{multline}
The large time asymptotics can be obtained by the saddle point approximation, i.e., by expanding the exponent to the second order around $\phi=0$ and substituting $\phi=0$ in the rest of the integrand, which yields the diffusive result
\be
\vev{\varrho_q(x,t)\varrho_q(0,0)} \approx 
\rho^2\frac{\exp\left[-\frac{(\pr-\pl)^2}{2(\pr+\pl)}\right]}{\sqrt{2\pi(\pr+\pl)}}
\approx\rho\frac{ e^{-x^2/(4Dt)}}{\sqrt{4\pi Dt}}\,.
\ee

The autocorrelation can be obtained in an analytic form. At $x=0,$ the $v$-integral in Eq. \erf{eq:qq} vanishes (the integrand is an odd function), and $\pr=\pl=t/(2\tau).$ Then
\be
\vev{\varrho_q(0,t)\varrho_q(0,0)} = \int_0^{2\pi}\frac{\ud\phi}{2\pi}e^{-2\sin^2(\phi/2)t/\tau}\\
\left[\frac\rho{t}f(0)+\rho^2\cos^2(\phi/2)\right]=
\frac{\rho}{t}f(0) e^{-t/\tau} I_0(t/\tau)
+\frac12\rho^2 e^{-t/\tau} [I_0(t/\tau)+I_1(t/\tau)]\,.
\ee

\bibliography{sGcorr_paper}

\begin{thebibliography}{41}%
\makeatletter
\providecommand \@ifxundefined [1]{%
 \@ifx{#1\undefined}
}%
\providecommand \@ifnum [1]{%
 \ifnum #1\expandafter \@firstoftwo
 \else \expandafter \@secondoftwo
 \fi
}%
\providecommand \@ifx [1]{%
 \ifx #1\expandafter \@firstoftwo
 \else \expandafter \@secondoftwo
 \fi
}%
\providecommand \natexlab [1]{#1}%
\providecommand \enquote  [1]{``#1''}%
\providecommand \bibnamefont  [1]{#1}%
\providecommand \bibfnamefont [1]{#1}%
\providecommand \citenamefont [1]{#1}%
\providecommand \href@noop [0]{\@secondoftwo}%
\providecommand \href [0]{\begingroup \@sanitize@url \@href}%
\providecommand \@href[1]{\@@startlink{#1}\@@href}%
\providecommand \@@href[1]{\endgroup#1\@@endlink}%
\providecommand \@sanitize@url [0]{\catcode `\\12\catcode `\$12\catcode
  `\&12\catcode `\#12\catcode `\^12\catcode `\_12\catcode `\%12\relax}%
\providecommand \@@startlink[1]{}%
\providecommand \@@endlink[0]{}%
\providecommand \url  [0]{\begingroup\@sanitize@url \@url }%
\providecommand \@url [1]{\endgroup\@href {#1}{\urlprefix }}%
\providecommand \urlprefix  [0]{URL }%
\providecommand \Eprint [0]{\href }%
\providecommand \doibase [0]{http://dx.doi.org/}%
\providecommand \selectlanguage [0]{\@gobble}%
\providecommand \bibinfo  [0]{\@secondoftwo}%
\providecommand \bibfield  [0]{\@secondoftwo}%
\providecommand \translation [1]{[#1]}%
\providecommand \BibitemOpen [0]{}%
\providecommand \bibitemStop [0]{}%
\providecommand \bibitemNoStop [0]{.\EOS\space}%
\providecommand \EOS [0]{\spacefactor3000\relax}%
\providecommand \BibitemShut  [1]{\csname bibitem#1\endcsname}%
\let\auto@bib@innerbib\@empty
\bibitem [{\citenamefont {Giamarchi}(2003)}]{Giamarchibook}%
  \BibitemOpen
  \bibfield  {author} {\bibinfo {author} {\bibfnamefont {T.}~\bibnamefont
  {Giamarchi}},\ }\href@noop {} {\emph {\bibinfo {title} {{Quantum Physics in
  One Dimension}}}}\ (\bibinfo  {publisher} {Oxford University Press},\
  \bibinfo {address} {Oxford},\ \bibinfo {year} {2003})\BibitemShut {NoStop}%
\bibitem [{\citenamefont {Gogolin}\ \emph {et~al.}(1998)\citenamefont
  {Gogolin}, \citenamefont {Nersesyan},\ and\ \citenamefont
  {Tsvelik}}]{Gogolin}%
  \BibitemOpen
  \bibfield  {author} {\bibinfo {author} {\bibnamefont {Gogolin}}, \bibinfo
  {author} {\bibnamefont {Nersesyan}}, \ and\ \bibinfo {author} {\bibnamefont
  {Tsvelik}},\ }\href@noop {} {\emph {\bibinfo {title} {{Bosonization Approach
  to Strongly Correlated Systems}}}}\ (\bibinfo  {publisher} {Cambridge
  University Press},\ \bibinfo {address} {Cambridge},\ \bibinfo {year}
  {1998})\BibitemShut {NoStop}%
\bibitem [{\citenamefont {Essler}\ and\ \citenamefont
  {Konik}(2005)}]{Essler2005}%
  \BibitemOpen
  \bibfield  {author} {\bibinfo {author} {\bibfnamefont {F.~H.~L.}\
  \bibnamefont {Essler}}\ and\ \bibinfo {author} {\bibfnamefont {R.~M.}\
  \bibnamefont {Konik}},\ }\href {http://arxiv.org/abs/cond-mat/0412421
  http://www.worldscientific.com/worldscibooks/10.1142/5621} {\bibfield
  {journal} {\bibinfo  {journal} {in From Fields to Strings: Circumnavigating
  Theoretical Physics}\ } (\bibinfo {year} {2005})},\ \Eprint
  {http://arxiv.org/abs/0412421} {arXiv:0412421 [cond-mat]} \BibitemShut
  {NoStop}%
\bibitem [{\citenamefont {B\"{u}chler}\ \emph {et~al.}(2003)\citenamefont
  {B\"{u}chler}, \citenamefont {Blatter},\ and\ \citenamefont
  {Zwerger}}]{Buchler2003}%
  \BibitemOpen
  \bibfield  {author} {\bibinfo {author} {\bibfnamefont {H.~P.}\ \bibnamefont
  {B\"{u}chler}}, \bibinfo {author} {\bibfnamefont {G.}~\bibnamefont
  {Blatter}}, \ and\ \bibinfo {author} {\bibfnamefont {W.}~\bibnamefont
  {Zwerger}},\ }\href {\doibase 10.1103/PhysRevLett.90.130401} {\bibfield
  {journal} {\bibinfo  {journal} {Physical review letters}\ }\textbf {\bibinfo
  {volume} {90}},\ \bibinfo {pages} {130401} (\bibinfo {year} {2003})},\
  \Eprint {http://arxiv.org/abs/0208391} {arXiv:0208391 [cond-mat]}
  \BibitemShut {NoStop}%
\bibitem [{\citenamefont {Cazalilla}(2003)}]{Cazalilla_bosonizing2003}%
  \BibitemOpen
  \bibfield  {author} {\bibinfo {author} {\bibfnamefont {M.~A.}\ \bibnamefont
  {Cazalilla}},\ }\href {\doibase 10.1088/0953-4075/37/7/051} {\bibfield
  {journal} {\bibinfo  {journal} {Journal of Physics B: Atomic, Molecular and
  Optical Physics}\ }\textbf {\bibinfo {volume} {37}},\ \bibinfo {pages} {39}
  (\bibinfo {year} {2003})},\ \Eprint {http://arxiv.org/abs/0307033}
  {arXiv:0307033 [cond-mat]} \BibitemShut {NoStop}%
\bibitem [{\citenamefont {Gritsev}\ \emph {et~al.}(2007)\citenamefont
  {Gritsev}, \citenamefont {Polkovnikov},\ and\ \citenamefont
  {Demler}}]{Gritsev2007a}%
  \BibitemOpen
  \bibfield  {author} {\bibinfo {author} {\bibfnamefont {V.}~\bibnamefont
  {Gritsev}}, \bibinfo {author} {\bibfnamefont {A.}~\bibnamefont
  {Polkovnikov}}, \ and\ \bibinfo {author} {\bibfnamefont {E.}~\bibnamefont
  {Demler}},\ }\href {\doibase 10.1103/PhysRevB.75.174511} {\bibfield
  {journal} {\bibinfo  {journal} {Physical Review B - Condensed Matter and
  Materials Physics}\ }\textbf {\bibinfo {volume} {75}},\ \bibinfo {pages}
  {174511} (\bibinfo {year} {2007})},\ \Eprint {http://arxiv.org/abs/0701421}
  {arXiv:0701421 [cond-mat]} \BibitemShut {NoStop}%
\bibitem [{\citenamefont {{Schweigler}}\ \emph {et~al.}(2017)\citenamefont
  {{Schweigler}}, \citenamefont {{Kasper}}, \citenamefont {{Erne}},
  \citenamefont {{Mazets}}, \citenamefont {{Rauer}}, \citenamefont
  {{Cataldini}}, \citenamefont {{Langen}}, \citenamefont {{Gasenzer}},
  \citenamefont {{Berges}},\ and\ \citenamefont
  {{Schmiedmayer}}}]{Schweigler2017}%
  \BibitemOpen
  \bibfield  {author} {\bibinfo {author} {\bibfnamefont {T.}~\bibnamefont
  {{Schweigler}}}, \bibinfo {author} {\bibfnamefont {V.}~\bibnamefont
  {{Kasper}}}, \bibinfo {author} {\bibfnamefont {S.}~\bibnamefont {{Erne}}},
  \bibinfo {author} {\bibfnamefont {I.}~\bibnamefont {{Mazets}}}, \bibinfo
  {author} {\bibfnamefont {B.}~\bibnamefont {{Rauer}}}, \bibinfo {author}
  {\bibfnamefont {F.}~\bibnamefont {{Cataldini}}}, \bibinfo {author}
  {\bibfnamefont {T.}~\bibnamefont {{Langen}}}, \bibinfo {author}
  {\bibfnamefont {T.}~\bibnamefont {{Gasenzer}}}, \bibinfo {author}
  {\bibfnamefont {J.}~\bibnamefont {{Berges}}}, \ and\ \bibinfo {author}
  {\bibfnamefont {J.}~\bibnamefont {{Schmiedmayer}}},\ }\href {\doibase
  10.1038/nature22310} {\bibfield  {journal} {\bibinfo  {journal} {Nature}\
  }\textbf {\bibinfo {volume} {545}},\ \bibinfo {pages} {323} (\bibinfo {year}
  {2017})}\BibitemShut {NoStop}%
\bibitem [{\citenamefont {Zamolodchikov}\ and\ \citenamefont
  {Zamolodchikov}(1979)}]{Zamolodchikov1979}%
  \BibitemOpen
  \bibfield  {author} {\bibinfo {author} {\bibfnamefont {A.~B.}\ \bibnamefont
  {Zamolodchikov}}\ and\ \bibinfo {author} {\bibfnamefont {A.~B.}\ \bibnamefont
  {Zamolodchikov}},\ }\href {\doibase 10.1016/0003-4916(79)90391-9} {\bibfield
  {journal} {\bibinfo  {journal} {Annals of Physics}\ }\textbf {\bibinfo
  {volume} {120}},\ \bibinfo {pages} {253} (\bibinfo {year}
  {1979})}\BibitemShut {NoStop}%
\bibitem [{\citenamefont {Mussardo}(2010)}]{giuseppebook}%
  \BibitemOpen
  \bibfield  {author} {\bibinfo {author} {\bibfnamefont {G.}~\bibnamefont
  {Mussardo}},\ }\href@noop {} {\emph {\bibinfo {title} {{Statistical Field
  Theory}}}}\ (\bibinfo  {publisher} {Oxford University Press},\ \bibinfo
  {address} {Oxford},\ \bibinfo {year} {2010})\BibitemShut {NoStop}%
\bibitem [{\citenamefont {Essler}\ and\ \citenamefont
  {Konik}(2009)}]{Essler2009}%
  \BibitemOpen
  \bibfield  {author} {\bibinfo {author} {\bibfnamefont {F.~H.~L.}\
  \bibnamefont {Essler}}\ and\ \bibinfo {author} {\bibfnamefont {R.~M.}\
  \bibnamefont {Konik}},\ }\href {\doibase 10.1088/1742-5468/2009/09/p09018}
  {\bibfield  {journal} {\bibinfo  {journal} {J. Stat. Mech.}\ }\textbf
  {\bibinfo {volume} {2009}},\ \bibinfo {pages} {P09018} (\bibinfo {year}
  {2009})}\BibitemShut {NoStop}%
\bibitem [{\citenamefont {Pozsgay}\ and\ \citenamefont
  {Tak{\'{a}}cs}(2010)}]{Takacs2010}%
  \BibitemOpen
  \bibfield  {author} {\bibinfo {author} {\bibfnamefont {B.}~\bibnamefont
  {Pozsgay}}\ and\ \bibinfo {author} {\bibfnamefont {G.}~\bibnamefont
  {Tak{\'{a}}cs}},\ }\href {\doibase 10.1088/1742-5468/2010/11/p11012}
  {\bibfield  {journal} {\bibinfo  {journal} {J. Stat. Mech.}\ }\textbf
  {\bibinfo {volume} {2010}},\ \bibinfo {pages} {P11012} (\bibinfo {year}
  {2010})}\BibitemShut {NoStop}%
\bibitem [{\citenamefont {Sz{\'{e}}cs{\'{e}}nyi}\ and\ \citenamefont
  {Tak{\'{a}}cs}(2012)}]{Szecsenyi2012}%
  \BibitemOpen
  \bibfield  {author} {\bibinfo {author} {\bibfnamefont {I.~M.}\ \bibnamefont
  {Sz{\'{e}}cs{\'{e}}nyi}}\ and\ \bibinfo {author} {\bibfnamefont
  {G.}~\bibnamefont {Tak{\'{a}}cs}},\ }\href {\doibase
  10.1088/1742-5468/2012/12/p12002} {\bibfield  {journal} {\bibinfo  {journal}
  {J. Stat. Mech.}\ }\textbf {\bibinfo {volume} {2012}},\ \bibinfo {pages}
  {P12002} (\bibinfo {year} {2012})}\BibitemShut {NoStop}%
\bibitem [{\citenamefont {Bertini}\ \emph {et~al.}(2016)\citenamefont
  {Bertini}, \citenamefont {Collura}, \citenamefont {{De Nardis}},\ and\
  \citenamefont {Fagotti}}]{Bertini2016}%
  \BibitemOpen
  \bibfield  {author} {\bibinfo {author} {\bibfnamefont {B.}~\bibnamefont
  {Bertini}}, \bibinfo {author} {\bibfnamefont {M.}~\bibnamefont {Collura}},
  \bibinfo {author} {\bibfnamefont {J.}~\bibnamefont {{De Nardis}}}, \ and\
  \bibinfo {author} {\bibfnamefont {M.}~\bibnamefont {Fagotti}},\ }\href
  {\doibase 10.1103/PhysRevLett.117.207201} {\bibfield  {journal} {\bibinfo
  {journal} {Phys. Rev. Lett.}\ }\textbf {\bibinfo {volume} {117}},\ \bibinfo
  {pages} {207201} (\bibinfo {year} {2016})},\ \Eprint
  {http://arxiv.org/abs/1605.09790} {arXiv:1605.09790} \BibitemShut {NoStop}%
\bibitem [{\citenamefont {Castro-Alvaredo}\ \emph {et~al.}(2016)\citenamefont
  {Castro-Alvaredo}, \citenamefont {Doyon},\ and\ \citenamefont
  {Yoshimura}}]{Castro-Alvaredo2016a}%
  \BibitemOpen
  \bibfield  {author} {\bibinfo {author} {\bibfnamefont {O.~A.}\ \bibnamefont
  {Castro-Alvaredo}}, \bibinfo {author} {\bibfnamefont {B.}~\bibnamefont
  {Doyon}}, \ and\ \bibinfo {author} {\bibfnamefont {T.}~\bibnamefont
  {Yoshimura}},\ }\href {\doibase 10.1103/PhysRevX.6.041065} {\bibfield
  {journal} {\bibinfo  {journal} {Phys. Rev. X}\ }\textbf {\bibinfo {volume}
  {6}},\ \bibinfo {pages} {041065} (\bibinfo {year} {2016})},\ \Eprint
  {http://arxiv.org/abs/1605.07331} {arXiv:1605.07331} \BibitemShut {NoStop}%
\bibitem [{\citenamefont {Doyon}(2020)}]{Doyon2019a}%
  \BibitemOpen
  \bibfield  {author} {\bibinfo {author} {\bibfnamefont {B.}~\bibnamefont
  {Doyon}},\ }\href {\doibase 10.21468/SciPostPhysLectNotes.18} {\bibfield
  {journal} {\bibinfo  {journal} {SciPost Phys. Lect. Notes}\ ,\ \bibinfo
  {pages} {18}} (\bibinfo {year} {2020})},\ \Eprint
  {http://arxiv.org/abs/1912.08496} {arXiv:1912.08496} \BibitemShut {NoStop}%
\bibitem [{\citenamefont {Prosen}\ and\ \citenamefont
  {Ilievski}(2013)}]{Prosen2013}%
  \BibitemOpen
  \bibfield  {author} {\bibinfo {author} {\bibfnamefont {T.}~\bibnamefont
  {Prosen}}\ and\ \bibinfo {author} {\bibfnamefont {E.}~\bibnamefont
  {Ilievski}},\ }\href {\doibase 10.1103/PhysRevLett.111.057203} {\bibfield
  {journal} {\bibinfo  {journal} {Phys. Rev. Lett.}\ }\textbf {\bibinfo
  {volume} {111}} (\bibinfo {year} {2013}),\ 10.1103/PhysRevLett.111.057203},\
  \Eprint {http://arxiv.org/abs/1306.4498} {arXiv:1306.4498} \BibitemShut
  {NoStop}%
\bibitem [{\citenamefont {Ilievski}\ \emph {et~al.}(2018)\citenamefont
  {Ilievski}, \citenamefont {{De Nardis}}, \citenamefont {Medenjak},\ and\
  \citenamefont {Prosen}}]{Ilievski2018}%
  \BibitemOpen
  \bibfield  {author} {\bibinfo {author} {\bibfnamefont {E.}~\bibnamefont
  {Ilievski}}, \bibinfo {author} {\bibfnamefont {J.}~\bibnamefont {{De
  Nardis}}}, \bibinfo {author} {\bibfnamefont {M.}~\bibnamefont {Medenjak}}, \
  and\ \bibinfo {author} {\bibfnamefont {T.}~\bibnamefont {Prosen}},\ }\href
  {\doibase 10.1103/PhysRevLett.121.230602} {\bibfield  {journal} {\bibinfo
  {journal} {Phys. Rev. Lett.}\ }\textbf {\bibinfo {volume} {121}},\ \bibinfo
  {pages} {230602} (\bibinfo {year} {2018})},\ \Eprint
  {http://arxiv.org/abs/1806.03288} {arXiv:1806.03288} \BibitemShut {NoStop}%
\bibitem [{\citenamefont {Ilievski}\ \emph {et~al.}(2021)\citenamefont
  {Ilievski}, \citenamefont {{De Nardis}}, \citenamefont {Gopalakrishnan},
  \citenamefont {Vasseur},\ and\ \citenamefont {Ware}}]{Ilievski2021}%
  \BibitemOpen
  \bibfield  {author} {\bibinfo {author} {\bibfnamefont {E.}~\bibnamefont
  {Ilievski}}, \bibinfo {author} {\bibfnamefont {J.}~\bibnamefont {{De
  Nardis}}}, \bibinfo {author} {\bibfnamefont {S.}~\bibnamefont
  {Gopalakrishnan}}, \bibinfo {author} {\bibfnamefont {R.}~\bibnamefont
  {Vasseur}}, \ and\ \bibinfo {author} {\bibfnamefont {B.}~\bibnamefont
  {Ware}},\ }\href {\doibase 10.1103/PhysRevX.11.031023} {\bibfield  {journal}
  {\bibinfo  {journal} {Phys. Rev. X}\ }\textbf {\bibinfo {volume} {11}},\
  \bibinfo {pages} {031023} (\bibinfo {year} {2021})},\ \Eprint
  {http://arxiv.org/abs/2009.08425} {arXiv:2009.08425} \BibitemShut {NoStop}%
\bibitem [{\citenamefont {Medenjak}\ \emph {et~al.}(2017)\citenamefont
  {Medenjak}, \citenamefont {Karrasch},\ and\ \citenamefont
  {Prosen}}]{Medenjak2017}%
  \BibitemOpen
  \bibfield  {author} {\bibinfo {author} {\bibfnamefont {M.}~\bibnamefont
  {Medenjak}}, \bibinfo {author} {\bibfnamefont {C.}~\bibnamefont {Karrasch}},
  \ and\ \bibinfo {author} {\bibfnamefont {T.}~\bibnamefont {Prosen}},\ }\href
  {\doibase 10.1103/PhysRevLett.119.080602} {\bibfield  {journal} {\bibinfo
  {journal} {Phys. Rev. Lett.}\ }\textbf {\bibinfo {volume} {119}},\ \bibinfo
  {pages} {080602} (\bibinfo {year} {2017})},\ \Eprint
  {http://arxiv.org/abs/1702.04677} {arXiv:1702.04677} \BibitemShut {NoStop}%
\bibitem [{\citenamefont {Bertini}\ \emph {et~al.}(2021)\citenamefont
  {Bertini}, \citenamefont {Heidrich-Meisner}, \citenamefont {Karrasch},
  \citenamefont {Prosen}, \citenamefont {Steinigeweg},\ and\ \citenamefont
  {{\v{Z}}nidari{\v{c}}}}]{Bertini2021}%
  \BibitemOpen
  \bibfield  {author} {\bibinfo {author} {\bibfnamefont {B.}~\bibnamefont
  {Bertini}}, \bibinfo {author} {\bibfnamefont {F.}~\bibnamefont
  {Heidrich-Meisner}}, \bibinfo {author} {\bibfnamefont {C.}~\bibnamefont
  {Karrasch}}, \bibinfo {author} {\bibfnamefont {T.}~\bibnamefont {Prosen}},
  \bibinfo {author} {\bibfnamefont {R.}~\bibnamefont {Steinigeweg}}, \ and\
  \bibinfo {author} {\bibfnamefont {M.}~\bibnamefont {{\v{Z}}nidari{\v{c}}}},\
  }\href {\doibase 10.1103/RevModPhys.93.025003} {\bibfield  {journal}
  {\bibinfo  {journal} {Rev. Mod. Phys.}\ }\textbf {\bibinfo {volume} {93}},\
  \bibinfo {pages} {025003} (\bibinfo {year} {2021})},\ \Eprint
  {http://arxiv.org/abs/2003.03334} {arXiv:2003.03334} \BibitemShut {NoStop}%
\bibitem [{\citenamefont {Doyon}(2018)}]{Doyon2017f}%
  \BibitemOpen
  \bibfield  {author} {\bibinfo {author} {\bibfnamefont {B.}~\bibnamefont
  {Doyon}},\ }\href {\doibase 10.21468/SciPostPhys.5.5.054} {\bibfield
  {journal} {\bibinfo  {journal} {SciPost Phys.}\ }\textbf {\bibinfo {volume}
  {5}},\ \bibinfo {pages} {054} (\bibinfo {year} {2018})},\ \Eprint
  {http://arxiv.org/abs/1711.04568} {arXiv:1711.04568} \BibitemShut {NoStop}%
\bibitem [{\citenamefont {Doyon}\ and\ \citenamefont
  {Myers}(2019)}]{Doyon2019b}%
  \BibitemOpen
  \bibfield  {author} {\bibinfo {author} {\bibfnamefont {B.}~\bibnamefont
  {Doyon}}\ and\ \bibinfo {author} {\bibfnamefont {J.}~\bibnamefont {Myers}},\
  }\href {\doibase 10.1007/s00023-019-00860-w} {\bibfield  {journal} {\bibinfo
  {journal} {Ann. Henri Poincar{\'{e}}}\ } (\bibinfo {year} {2019}),\
  10.1007/s00023-019-00860-w},\ \Eprint {http://arxiv.org/abs/1902.00320}
  {arXiv:1902.00320} \BibitemShut {NoStop}%
\bibitem [{\citenamefont {Sachdev}\ and\ \citenamefont
  {Young}(1997)}]{Sachdev1996}%
  \BibitemOpen
  \bibfield  {author} {\bibinfo {author} {\bibfnamefont {S.}~\bibnamefont
  {Sachdev}}\ and\ \bibinfo {author} {\bibfnamefont {A.~P.}\ \bibnamefont
  {Young}},\ }\href {\doibase 10.1103/PhysRevLett.78.2220} {\bibfield
  {journal} {\bibinfo  {journal} {Physical Review Letters}\ }\textbf {\bibinfo
  {volume} {78}},\ \bibinfo {pages} {2220} (\bibinfo {year} {1997})},\ \Eprint
  {http://arxiv.org/abs/9609185} {arXiv:9609185 [cond-mat]} \BibitemShut
  {NoStop}%
\bibitem [{\citenamefont {Sachdev}\ and\ \citenamefont
  {Damle}(1997)}]{Sachdev1997}%
  \BibitemOpen
  \bibfield  {author} {\bibinfo {author} {\bibfnamefont {S.}~\bibnamefont
  {Sachdev}}\ and\ \bibinfo {author} {\bibfnamefont {K.}~\bibnamefont
  {Damle}},\ }\href {\doibase 10.1103/PhysRevLett.78.943} {\bibfield  {journal}
  {\bibinfo  {journal} {Physical Review Letters}\ }\textbf {\bibinfo {volume}
  {78}},\ \bibinfo {pages} {943} (\bibinfo {year} {1997})},\ \Eprint
  {http://arxiv.org/abs/9610115} {arXiv:9610115 [cond-mat]} \BibitemShut
  {NoStop}%
\bibitem [{\citenamefont {Damle}\ and\ \citenamefont
  {Sachdev}(2005)}]{Damle2005}%
  \BibitemOpen
  \bibfield  {author} {\bibinfo {author} {\bibfnamefont {K.}~\bibnamefont
  {Damle}}\ and\ \bibinfo {author} {\bibfnamefont {S.}~\bibnamefont
  {Sachdev}},\ }\href {\doibase 10.1103/PhysRevLett.95.187201} {\bibfield
  {journal} {\bibinfo  {journal} {Physical Review Letters}\ }\textbf {\bibinfo
  {volume} {95}},\ \bibinfo {pages} {187201} (\bibinfo {year} {2005})},\
  \Eprint {http://arxiv.org/abs/0507380} {arXiv:0507380 [cond-mat]}
  \BibitemShut {NoStop}%
\bibitem [{\citenamefont {Rapp}\ and\ \citenamefont
  {Zar\'{a}nd}(2006)}]{Rapp2006}%
  \BibitemOpen
  \bibfield  {author} {\bibinfo {author} {\bibfnamefont {A.}~\bibnamefont
  {Rapp}}\ and\ \bibinfo {author} {\bibfnamefont {G.}~\bibnamefont
  {Zar\'{a}nd}},\ }\href {\doibase 10.1103/PhysRevB.74.014433} {\bibfield
  {journal} {\bibinfo  {journal} {Physical Review B}\ }\textbf {\bibinfo
  {volume} {74}},\ \bibinfo {pages} {014433} (\bibinfo {year} {2006})},\
  \Eprint {http://arxiv.org/abs/0507390} {arXiv:0507390 [cond-mat]}
  \BibitemShut {NoStop}%
\bibitem [{\citenamefont {Evangelisti}(2013)}]{Evangelisti2013}%
  \BibitemOpen
  \bibfield  {author} {\bibinfo {author} {\bibfnamefont {S.}~\bibnamefont
  {Evangelisti}},\ }\href {\doibase 10.1088/1742-5468/2013/04/P04003}
  {\bibfield  {journal} {\bibinfo  {journal} {Journal of Statistical Mechanics:
  Theory and Experiment}\ ,\ \bibinfo {pages} {P04003}} (\bibinfo {year}
  {2013})},\ \Eprint {http://arxiv.org/abs/1210.4028} {arXiv:1210.4028}
  \BibitemShut {NoStop}%
\bibitem [{\citenamefont {Kormos}\ and\ \citenamefont
  {Zar{\'{a}}nd}(2016)}]{Kormos2015}%
  \BibitemOpen
  \bibfield  {author} {\bibinfo {author} {\bibfnamefont {M.}~\bibnamefont
  {Kormos}}\ and\ \bibinfo {author} {\bibfnamefont {G.}~\bibnamefont
  {Zar{\'{a}}nd}},\ }\href {\doibase 10.1103/PhysRevE.93.062101} {\bibfield
  {journal} {\bibinfo  {journal} {Phys. Rev. E}\ }\textbf {\bibinfo {volume}
  {93}},\ \bibinfo {pages} {062101} (\bibinfo {year} {2016})},\ \Eprint
  {http://arxiv.org/abs/1507.02708} {arXiv:1507.02708} \BibitemShut {NoStop}%
\bibitem [{\citenamefont {Bertini}\ \emph {et~al.}(2019)\citenamefont
  {Bertini}, \citenamefont {Piroli},\ and\ \citenamefont
  {Kormos}}]{Bertini2019}%
  \BibitemOpen
  \bibfield  {author} {\bibinfo {author} {\bibfnamefont {B.}~\bibnamefont
  {Bertini}}, \bibinfo {author} {\bibfnamefont {L.}~\bibnamefont {Piroli}}, \
  and\ \bibinfo {author} {\bibfnamefont {M.}~\bibnamefont {Kormos}},\ }\href
  {\doibase 10.1103/PhysRevB.100.035108} {\bibfield  {journal} {\bibinfo
  {journal} {Phys. Rev. B}\ }\textbf {\bibinfo {volume} {100}},\ \bibinfo
  {pages} {035108} (\bibinfo {year} {2019})},\ \Eprint
  {http://arxiv.org/abs/1904.02696} {arXiv:1904.02696} \BibitemShut {NoStop}%
\bibitem [{\citenamefont {Moca}\ \emph {et~al.}(2017)\citenamefont {Moca},
  \citenamefont {Kormos},\ and\ \citenamefont {Zar{\'{a}}nd}}]{Moca2016}%
  \BibitemOpen
  \bibfield  {author} {\bibinfo {author} {\bibfnamefont {C.~P.}\ \bibnamefont
  {Moca}}, \bibinfo {author} {\bibfnamefont {M.}~\bibnamefont {Kormos}}, \ and\
  \bibinfo {author} {\bibfnamefont {G.}~\bibnamefont {Zar{\'{a}}nd}},\ }\href
  {\doibase 10.1103/PhysRevLett.119.100603} {\bibfield  {journal} {\bibinfo
  {journal} {Phys. Rev. Lett.}\ }\textbf {\bibinfo {volume} {119}},\ \bibinfo
  {pages} {100603} (\bibinfo {year} {2017})},\ \Eprint
  {http://arxiv.org/abs/1609.00974} {arXiv:1609.00974} \BibitemShut {NoStop}%
\bibitem [{\citenamefont {Werner}\ \emph {et~al.}(2019)\citenamefont {Werner},
  \citenamefont {Moca}, \citenamefont {Legeza}, \citenamefont {Kormos},\ and\
  \citenamefont {Zar{\'{a}}nd}}]{Werner2019}%
  \BibitemOpen
  \bibfield  {author} {\bibinfo {author} {\bibfnamefont {M.~A.}\ \bibnamefont
  {Werner}}, \bibinfo {author} {\bibfnamefont {C.~P.}\ \bibnamefont {Moca}},
  \bibinfo {author} {\bibfnamefont {{\"{O}}.}~\bibnamefont {Legeza}}, \bibinfo
  {author} {\bibfnamefont {M.}~\bibnamefont {Kormos}}, \ and\ \bibinfo {author}
  {\bibfnamefont {G.}~\bibnamefont {Zar{\'{a}}nd}},\ }\href {\doibase
  10.1103/PhysRevB.100.035401} {\bibfield  {journal} {\bibinfo  {journal}
  {Phys. Rev. B}\ }\textbf {\bibinfo {volume} {100}},\ \bibinfo {pages} {1}
  (\bibinfo {year} {2019})},\ \Eprint {http://arxiv.org/abs/1902.08587}
  {arXiv:1902.08587} \BibitemShut {NoStop}%
\bibitem [{\citenamefont {Zamolodchikov}(1995)}]{Zamolodchikov1995}%
  \BibitemOpen
  \bibfield  {author} {\bibinfo {author} {\bibfnamefont {A.~B.}\ \bibnamefont
  {Zamolodchikov}},\ }\href {\doibase 10.1142/S0217751X9500053X} {\bibfield
  {journal} {\bibinfo  {journal} {Int. J. Mod. Phys. A}\ }\textbf {\bibinfo
  {volume} {10}},\ \bibinfo {pages} {1125} (\bibinfo {year}
  {1995})}\BibitemShut {NoStop}%
\bibitem [{\citenamefont {Luther}\ and\ \citenamefont
  {Emery}(1974)}]{Luther-Emery1974}%
  \BibitemOpen
  \bibfield  {author} {\bibinfo {author} {\bibfnamefont {A.}~\bibnamefont
  {Luther}}\ and\ \bibinfo {author} {\bibfnamefont {V.~J.}\ \bibnamefont
  {Emery}},\ }\href {\doibase 10.1103/PhysRevLett.33.589} {\bibfield  {journal}
  {\bibinfo  {journal} {Physical Review Letters}\ }\textbf {\bibinfo {volume}
  {33}},\ \bibinfo {pages} {589} (\bibinfo {year} {1974})}\BibitemShut
  {NoStop}%
\bibitem [{\citenamefont {Coleman}(1975)}]{Coleman1975}%
  \BibitemOpen
  \bibfield  {author} {\bibinfo {author} {\bibfnamefont {S.}~\bibnamefont
  {Coleman}},\ }\href {\doibase 10.1103/PhysRevD.11.2088} {\bibfield  {journal}
  {\bibinfo  {journal} {Physical Review D}\ }\textbf {\bibinfo {volume} {11}},\
  \bibinfo {pages} {2088} (\bibinfo {year} {1975})}\BibitemShut {NoStop}%
\bibitem [{\citenamefont {Altshuler}\ \emph {et~al.}(2006)\citenamefont
  {Altshuler}, \citenamefont {Konik},\ and\ \citenamefont
  {Tsvelik}}]{Altshuler2006}%
  \BibitemOpen
  \bibfield  {author} {\bibinfo {author} {\bibfnamefont {B.}~\bibnamefont
  {Altshuler}}, \bibinfo {author} {\bibfnamefont {R.}~\bibnamefont {Konik}}, \
  and\ \bibinfo {author} {\bibfnamefont {A.}~\bibnamefont {Tsvelik}},\ }\href
  {\doibase 10.1016/j.nuclphysb.2006.01.022} {\bibfield  {journal} {\bibinfo
  {journal} {Nuclear Physics B}\ }\textbf {\bibinfo {volume} {739}},\ \bibinfo
  {pages} {311} (\bibinfo {year} {2006})}\BibitemShut {NoStop}%
\bibitem [{\citenamefont {Imamura}\ \emph {et~al.}(2021)\citenamefont
  {Imamura}, \citenamefont {Mallick},\ and\ \citenamefont
  {Sasamoto}}]{Imamura2021}%
  \BibitemOpen
  \bibfield  {author} {\bibinfo {author} {\bibfnamefont {T.}~\bibnamefont
  {Imamura}}, \bibinfo {author} {\bibfnamefont {K.}~\bibnamefont {Mallick}}, \
  and\ \bibinfo {author} {\bibfnamefont {T.}~\bibnamefont {Sasamoto}},\ }\href
  {https://arxiv.org/pdf/1810.06131.pdf http://arxiv.org/abs/1810.06131}
  {\bibfield  {journal} {\bibinfo  {journal} {Commun. Math. Phys.}\ }\textbf
  {\bibinfo {volume} {384}},\ \bibinfo {pages} {1409} (\bibinfo {year}
  {2021})},\ \Eprint {http://arxiv.org/abs/1810.06131} {arXiv:1810.06131}
  \BibitemShut {NoStop}%
\bibitem [{\citenamefont {Krajnik}\ \emph
  {et~al.}(2022{\natexlab{a}})\citenamefont {Krajnik}, \citenamefont {Schmidt},
  \citenamefont {Pasquier}, \citenamefont {Ilievski},\ and\ \citenamefont
  {Prosen}}]{Krajnik2022}%
  \BibitemOpen
  \bibfield  {author} {\bibinfo {author} {\bibfnamefont {{\v{Z}}.}~\bibnamefont
  {Krajnik}}, \bibinfo {author} {\bibfnamefont {J.}~\bibnamefont {Schmidt}},
  \bibinfo {author} {\bibfnamefont {V.}~\bibnamefont {Pasquier}}, \bibinfo
  {author} {\bibfnamefont {E.}~\bibnamefont {Ilievski}}, \ and\ \bibinfo
  {author} {\bibfnamefont {T.}~\bibnamefont {Prosen}},\ }\href {\doibase
  10.1103/PhysRevLett.128.160601} {\bibfield  {journal} {\bibinfo  {journal}
  {Phys. Rev. Lett.}\ }\textbf {\bibinfo {volume} {128}},\ \bibinfo {pages}
  {160601} (\bibinfo {year} {2022}{\natexlab{a}})},\ \Eprint
  {http://arxiv.org/abs/2201.05126} {arXiv:2201.05126} \BibitemShut {NoStop}%
\bibitem [{\citenamefont {Krajnik}\ \emph
  {et~al.}(2022{\natexlab{b}})\citenamefont {Krajnik}, \citenamefont {Schmidt},
  \citenamefont {Pasquier}, \citenamefont {Prosen},\ and\ \citenamefont
  {Ilievski}}]{Krajnik2022c}%
  \BibitemOpen
  \bibfield  {author} {\bibinfo {author} {\bibfnamefont {{\v{Z}}.}~\bibnamefont
  {Krajnik}}, \bibinfo {author} {\bibfnamefont {J.}~\bibnamefont {Schmidt}},
  \bibinfo {author} {\bibfnamefont {V.}~\bibnamefont {Pasquier}}, \bibinfo
  {author} {\bibfnamefont {T.}~\bibnamefont {Prosen}}, \ and\ \bibinfo {author}
  {\bibfnamefont {E.}~\bibnamefont {Ilievski}},\ }\href
  {https://arxiv.org/abs/2208.01463} {\  (\bibinfo {year}
  {2022}{\natexlab{b}})},\ \Eprint {http://arxiv.org/abs/2208.01463}
  {arXiv:2208.01463} \BibitemShut {NoStop}%
\bibitem [{\citenamefont {Krajnik}\ \emph
  {et~al.}(2022{\natexlab{c}})\citenamefont {Krajnik}, \citenamefont
  {Ilievski},\ and\ \citenamefont {Prosen}}]{Krajnik2022b}%
  \BibitemOpen
  \bibfield  {author} {\bibinfo {author} {\bibfnamefont {{\v{Z}}.}~\bibnamefont
  {Krajnik}}, \bibinfo {author} {\bibfnamefont {E.}~\bibnamefont {Ilievski}}, \
  and\ \bibinfo {author} {\bibfnamefont {T.}~\bibnamefont {Prosen}},\ }\href
  {\doibase 10.1103/PhysRevLett.128.090604} {\bibfield  {journal} {\bibinfo
  {journal} {Phys. Rev. Lett.}\ }\textbf {\bibinfo {volume} {128}},\ \bibinfo
  {pages} {090604} (\bibinfo {year} {2022}{\natexlab{c}})},\ \Eprint
  {http://arxiv.org/abs/2109.13088} {arXiv:2109.13088} \BibitemShut {NoStop}%
\bibitem [{\citenamefont {Gopalakrishnan}\ \emph {et~al.}(2022)\citenamefont
  {Gopalakrishnan}, \citenamefont {Morningstar}, \citenamefont {Vasseur},\ and\
  \citenamefont {Khemani}}]{Gopalakrishnan2022}%
  \BibitemOpen
  \bibfield  {author} {\bibinfo {author} {\bibfnamefont {S.}~\bibnamefont
  {Gopalakrishnan}}, \bibinfo {author} {\bibfnamefont {A.}~\bibnamefont
  {Morningstar}}, \bibinfo {author} {\bibfnamefont {R.}~\bibnamefont
  {Vasseur}}, \ and\ \bibinfo {author} {\bibfnamefont {V.}~\bibnamefont
  {Khemani}},\ }\href {\doibase 10.48550/arxiv.2203.09526} {\  (\bibinfo {year}
  {2022}),\ 10.48550/arxiv.2203.09526},\ \Eprint
  {http://arxiv.org/abs/2203.09526} {arXiv:2203.09526} \BibitemShut {NoStop}%
\bibitem [{\citenamefont {Del Vecchio Del~Vecchio}\ \emph
  {et~al.}()\citenamefont {Del Vecchio Del~Vecchio}, \citenamefont
  {Bastianello}, \citenamefont {Kormos},\ and\ \citenamefont {Doyon}}]{inprep}%
  \BibitemOpen
  \bibfield  {author} {\bibinfo {author} {\bibfnamefont {G.}~\bibnamefont {Del
  Vecchio Del~Vecchio}}, \bibinfo {author} {\bibfnamefont {A.}~\bibnamefont
  {Bastianello}}, \bibinfo {author} {\bibfnamefont {M.}~\bibnamefont {Kormos}},
  \ and\ \bibinfo {author} {\bibfnamefont {B.}~\bibnamefont {Doyon}},\
  }\href@noop {} {}\bibinfo {note} {\emph{in preparation}}\BibitemShut
  {NoStop}%
\end{thebibliography}%

\end{document}